\documentclass[conference]{IEEEtran}
\usepackage{cite}

\usepackage{amssymb}
\usepackage{booktabs}
\usepackage{listings}
\usepackage{xcolor} 
\usepackage{amsmath}
\usepackage[most]{tcolorbox}
\tcbuselibrary{breakable}
\usepackage[hidelinks]{hyperref}
\usepackage{multirow} 
\usepackage{colortbl}
\usepackage[utf8]{inputenc}
\usepackage{CJKutf8}
\usepackage{pifont}
\usepackage{float}
\usepackage{multicol}
\usepackage{threeparttable}
\usepackage{tikz}
\usetikzlibrary{shapes.geometric, arrows.meta, positioning, backgrounds, calc}
\usepackage{subcaption}

\usepackage{hyperref}
\hypersetup{
  colorlinks = true,
  linkcolor = purple,
  anchorcolor = purple,
  citecolor = purple,
  filecolor = purple,
  urlcolor = purple
}
\usepackage{xurl}
\usepackage[hyphenbreaks]{breakurl}

\definecolor{js-background}{RGB}{248,248,248}
\definecolor{js-comment}{RGB}{63,127,95}
\definecolor{js-keyword}{RGB}{123,100,255}
\definecolor{js-string}{RGB}{163,21,21}
\definecolor{js-identifier}{RGB}{0,0,0}

\lstdefinestyle{JavaScript}{
    backgroundcolor=\color{js-background},
    basicstyle=\ttfamily\small,
    breaklines=true,
    breakatwhitespace=false,
    frame=single,
    captionpos=b,
    commentstyle=\color{js-comment},
    keywordstyle=\color{js-keyword},
    stringstyle=\color{js-string},
    identifierstyle=\color{js-identifier},
    morekeywords={var, new, function, document, eval},
    morecomment=[l]{//},
    tabsize=2,
    showstringspaces=false,
    columns=flexible
}

\PassOptionsToPackage{pdftex}{graphicx}
\usepackage{graphicx}

\usepackage{xspace}
\newcommand{\OurTool}{{\sc JSimplifier}\xspace}

\usepackage[available, functional, reproduced]{ndssbadges}
\begin{document}
\begin{CJK*}{UTF8}{gbsn}

%
\title{From Obfuscated to Obvious:\\
A Comprehensive JavaScript Deobfuscation\\
Tool for Security Analysis}

	

%
\author{\IEEEauthorblockN{Dongchao Zhou\textsuperscript{1,2},
Lingyun Ying\textsuperscript{2,*}\thanks{* Corresponding authors.},
Huajun Chai\textsuperscript{2} and
Dongbin Wang\textsuperscript{1,*}\thanks{Dongbin Wang is also with the Engineering Research Center of Blockchain and Network Convergence Technology, Ministry of Education, and the National Engineering Laboratory for Mobile Networks, Beijing University of Posts and Telecommunications (BUPT).}} 
\IEEEauthorblockA{\textsuperscript{1}Beijing University of Post and Telecommunication \{zdc, dbwang\}@bupt.edu.cn}
\IEEEauthorblockA{\textsuperscript{2}QI-ANXIN Technology Research Institute \{yinglingyun, chaihuajun\}@qianxin.com}
}



\IEEEoverridecommandlockouts
\makeatletter\def\@IEEEpubidpullup{6.5\baselineskip}\makeatother
\IEEEpubid{\parbox{\columnwidth}{
		Network and Distributed System Security (NDSS) Symposium 2026\\
		23-27 February 2026, San Diego, CA, USA\\
		ISBN 979-8-9919276-8-0\\
		https://dx.doi.org/10.14722/ndss.2026.242198.\\
		www.ndss-symposium.org
}
\hspace{\columnsep}\makebox[\columnwidth]{}}

\maketitle
\renewcommand{\thefootnote}{\fnsymbol{footnote}} 
\footnotetext[1]{This document corresponds to the full version of our NDSS 2026 paper.}
\begin{abstract}
JavaScript's widespread adoption has made it an attractive target for malicious attackers who employ sophisticated obfuscation techniques to conceal harmful code. Current deobfuscation tools suffer from critical limitations that severely restrict their practical effectiveness. Existing tools struggle with diverse input formats, address only specific obfuscation types, and produce cryptic output that impedes human analysis. 

To address these challenges, we present \OurTool, a comprehensive deobfuscation tool using a multi-stage pipeline with preprocessing, abstract syntax tree-based static analysis, dynamic execution tracing, and Large Language Model (LLM)-enhanced identifier renaming. We also introduce multi-dimensional evaluation metrics that integrate control/data flow analysis, code simplification assessment, entropy measures and LLM-based readability assessments. 

We construct and release the largest real-world obfuscated JavaScript dataset with 44,421 samples (23,212 wild malicious + 21,209 benign samples). Evaluation shows \OurTool outperforms existing tools with 100\% processing capability across 20 obfuscation techniques, 100\% correctness on evaluation subsets, 88.2\% code complexity reduction, and over 4-fold readability improvement validated by multiple LLMs. Our results advance benchmarks for JavaScript deobfuscation research and practical security applications.
\end{abstract}


%

\section{Introduction}
JavaScript's dynamic nature (such as dynamic typing, runtime code modification, and code generation) and widespread adoption have made it a prime target for malicious exploitation. Attackers implement complex obfuscation techniques that evade traditional static analysis~\cite{ren2023empirical,ndichu2020deobfuscation,gong2018dynamic,vilchik2024static}. 
Real-world campaigns (e.g., SocGholish~\cite{trendmicro:socgholish2025}) employ multi-layered techniques, such as string encoding, function name scrambling, and anti-analysis evasion methods~\cite{trendmicro:socgholish2025,proofpoint2022socgholish}, to obfuscate JavaScript code. The obfuscated code presents significant barriers to security analysis, making it extremely difficult for analysts to understand malicious behavior and develop effective countermeasures~\cite{brezinski2023metamorphic}.

To comprehensively understand the current deobfuscation landscape, we conducted a literature review and sample analysis to identify and categorize JavaScript obfuscation techniques. Based on the fundamental aspects of code they modify, we categorized 20 distinct obfuscation techniques across four main categories: (1) Lexical-level obfuscation (5 techniques) that transforms basic lexical elements like identifiers and literals while preserving syntactic structure; (2) Syntactic-level obfuscation (6 techniques) that modifies structural organization through advanced encoding methods and functional transformations; (3) Semantic-level obfuscation (7 techniques) that fundamentally alters implementation logic through string arrays, control flow manipulation, and dynamic code generation; and (4) Multi-layer obfuscation (2 techniques) that combines multiple strategies from different levels to create highly complex transformations. 

Current JavaScript deobfuscation tools, such as JStillery~\cite{jstillery}, illuminateJS~\cite{illuminatejs}, and JSNice~\cite{raychev2015predicting}, face three critical limitations when handling these diverse obfuscation techniques~\cite{kasada2023javascript}. 
First, they struggle with diverse input formats, crashing on malformed syntax and mixed encodings. Second, they focus on specific patterns and lack multi-layer support. Static analysis tools (e.g., JSNice~\cite{raychev2015predicting}, JSbeautifier.org~\cite{lielmanis2024jsbeautify}) fail on runtime-dependent obfuscation such as dynamic code generation and eval-based transformations, while dynamic analysis approaches (e.g., synchrony~\cite{synchrony2024}, JS-deobfuscator~\cite{bensb_deobfuscator_io}) struggle with control flow manipulations and string array deobfuscation. 
Third, they produce cryptic identifiers that impede human analysis.

To address these limitations, we propose a comprehensive deobfuscation tool, \OurTool, which uses a multi-stage pipeline. First, we develop robust code normalization through fault-tolerant parsing, character encoding standardization, and bundler pattern unwrapping, enabling reliable processing of diverse malformed inputs that crash existing tools.  
Second, our deobfuscation engine adopts a dual-strategy approach that combines static Abstract Syntax Tree (AST) analysis with controlled execution monitoring to address multi-layer obfuscation complexity. Unlike existing static-only tools that fail on runtime-dependent transformations, our enhanced static analysis handles complex control flow manipulations and string array deobfuscation through comprehensive expression evaluation and intelligent scope management. Unlike existing dynamic-only approaches that struggle with static patterns, our controlled execution monitoring selectively executes obfuscated code segments within sandboxed environments and integrates results back into static analysis, enabling comprehensive support for multi-layer transformations.
Finally, we apply Large Language Model (LLM)-based semantic enhancement for context-aware identifier renaming, transforming mechanically correct but cryptic output into human-readable code suitable for security analysis workflows.

For fairly and comprehensively evaluate our approach, we curated two dedicated datasets:
a) \textit{MalJS}, a large-scale dataset of 23,212 wild malicious JavaScript samples curated from over 10 million ones, covering all known obfuscation techniques. This dataset provides real-world samples with diverse and multi-layer obfuscation techniques, unlike existing datasets with artificial samples. 
b) \textit{BenignJS} comprises 2,000 manually verified benign samples from GitHub and web crawling. 
Moreover, we also adopt \textit{CombiBench}, a complex combination obfuscation subset from JsDeObsBench~\cite{jsdeobsbench2025}, for a fair comparison with existing methods.

Our evaluation methodology encompasses six complementary dimensions to comprehensively assess deobfuscation performance. First, we conduct a deobfuscation capability evaluation across 20 categorized obfuscation techniques. Second, we assess the correctness of deobfuscated code. Third, we evaluate semantic consistency to the original code. Fourth, we evaluate code simplification using complexity metrics from JsDeObsBench. Fifth, we validate effectiveness through entropy-based analysis. Sixth, we conduct multi-LLM readability assessment.

We compare \OurTool with 12 existing approaches—including nine traditional tools (JSNice~\cite{raychev2015predicting}, JStillery~\cite{jstillery}, illuminateJS~\cite{illuminatejs}, etc.) and three LLM-based solutions (GPT-o1, Gemini-2.0-Flash, DeepSeek-R1)—selecting appropriate baselines for each evaluation dimension. The results demonstrate strong performance of \OurTool across multiple evaluation dimensions. 
Specifically, our approach demonstrates comprehensive success across all 20 obfuscation techniques with 100\% processing capability and correctness on a subset of BenignJS with artificially applied obfuscation. The results maintain semantic consistency with 93.78\% Control Flow Graph (CFG) similarity scores and 95.84\% Data Dependence Graph (DDG) preservation rates. 
In addition, our approach achieves a 0.8820 Halstead length reduction in code simplification evaluation on CombiBench, representing an 88.20\% reduction in code complexity. This outperforms traditional deobfuscation approaches: JS-deobfuscator~\cite{bensb_deobfuscator_io} scores 0.0015, and Synchrony~\cite{synchrony2024} scores 0.7889.

Moreover, large-scale validation on MalJS demonstrates a comprehensive entropy reduction, while independent LLM-based readability assessment confirms an average improvement of 466.94\% across four evaluation models (Claude-3.7-Sonnet, Gemini-2.5-Pro, DeepSeek-R1, GPT-o3). \OurTool transforms unreadable obfuscated code into comprehensible versions suitable for security analysis workflows. 
Furthermore, detailed case studies, including analysis of the JSFireTruck~\cite{unit42_jsfiretruck} campaign, demonstrate \OurTool's practical effectiveness in real-world threat analysis contexts where traditional tools fail to provide adequate support.

We summarize our main contributions as follows:
\begin{itemize}
    \item We present \OurTool, a comprehensive JavaScript deobfuscation tool that integrates code normalization, dual-strategy deobfuscation (static AST analysis and controlled execution), and LLM-based semantic enhancement for robust and readable code recovery.
    \item We demonstrate \OurTool's strong performance with 100\% processing capability across all obfuscation techniques, exceptional code simplification (0.8820 Halstead reduction), and significant readability improvements validated by multiple LLMs.
    \item We construct and release the largest real-world obfuscated JavaScript dataset comprising 44,421 samples. This dataset provides valuable insights into real-world obfuscation practices and enables more effective development and evaluation of deobfuscation methods.
\end{itemize}

\textbf{Open Source.}
\OurTool and datasets are publicly available at \href{https://doi.org/10.5281/zenodo.17531661}{Zenodo} and GitHub~\cite{jsimplifier}. Detailed usage instructions are in Appendix~\ref{appendix:artifact}.
\label{intro}

\section{background}
\subsection{JavaScript}
JavaScript is a widely-used programming language that powers web applications and has expanded to server-side and mobile development. Its dynamic nature and flexible syntax make it attractive to malicious attackers for code obfuscation purposes.
The language's dynamic features create inherent security vulnerabilities that facilitate sophisticated obfuscation techniques. JavaScript allows variables to change types during execution, enabling transformations where \texttt{var x = 5} can later become \texttt{x = "hello"}. Additionally, the language supports runtime code modification through functions like \texttt{eval("alert('Hello')")} or dynamic property changes such as \texttt{window['alert'] = function()\{...\}}, providing attackers with powerful tools for code manipulation. 

Most critically, JavaScript's loose type system provides powerful obfuscation opportunities through its automatic but inconsistent type conversion rules. The \texttt{+} operator performs string concatenation if either operand is a string (\texttt{"5" + 3} becomes \texttt{"53"}), while other operators like \texttt{-} always perform numeric operations (\texttt{"5" - 3} results in \texttt{2}). These conversion rules enable complex obfuscation chains where expressions like \texttt{+!![]} equal 1: the empty array \texttt{[]} is truthy~\cite{ecmascript2023}, \texttt{!![]} becomes boolean \texttt{True}, and \texttt{+} converts the boolean value to number 1. In contrast, \texttt{!![] + []} equals \texttt{"true"} because when added to an empty array, JavaScript converts both operands to strings for concatenation. 

Attackers exploit these multi-step conversions to hide simple values behind complex expressions. This enables advanced techniques like JSFUCK~\cite{jsf*ck:handwiki}, which uses only six characters \texttt{[]()!+} to encode arbitrary JavaScript logic.
Furthermore, JavaScript treats functions as values that can be stored in variables, passed as arguments, and created dynamically. This enables attackers to build complex control flows and use string-based code execution through \texttt{eval()} to create layers of encoding that effectively hide malicious intent from static analysis tools.

\subsection{JavaScript Obfuscation}
\label{sec:js_obfuscation}
We systematically examined JavaScript obfuscation techniques through comprehensive analysis of both academic literature and practical implementations. Our investigation included 12 mainstream obfuscation tools and a literature survey on existing academic research on JavaScript obfuscation techniques~\cite{skolka2019anything, xu2013power, dsn21, imc20, jack2023, hidenoseek, wobfuscator, ieeeaccess19, caai20}. Additionally, we collected and analyzed real-world obfuscated JavaScript samples from malware datasets (VirusTotal) and commercial obfuscators (Jscrambler~\cite{jscrambler}, Jshaman~\cite{jshaman}).

Based on this comprehensive effort, we categorize JavaScript obfuscation techniques according to the fundamental aspects of code they modify. This taxonomy encompasses four main categories covering 20 distinct obfuscation techniques, as detailed in Appendix~\ref{appendix:obfuscator_options}. 

\noindent {\bf Lexical-level Obfuscation.} These techniques transform basic lexical elements while preserving underlying syntactic and semantic structure. Key techniques include identifier renaming (T0), which replaces meaningful variable and function names with random strings like \texttt{\_0x62ab7d}; indirect property access (T1), which converts \texttt{console.log} to \texttt{console['\textbackslash x6c\textbackslash x6f\textbackslash x67']}; arithmetization (T2), which replaces simple numbers with complex expressions like \texttt{683517 \textasciicircum{} 683398} for 123; string encoding (T3), which encodes strings using Base64 or other schemes; and boolean encoding (T4), which transforms \texttt{True} into \texttt{!![]}. 

\noindent {\bf Syntactic-level Obfuscation.} These techniques modify structural organization without changing fundamental logic. This includes assignment-to-function transformations (T6), which wraps simple assignments in immediately invoked function expressions; string reversal (T7), which stores strings backwards and reconstructs them using \texttt{split}, \texttt{reverse}, and \texttt{join}; and advanced encoding methods such as AAEncode (T8), which uses special Unicode characters to encode JavaScript; JJEncode (T9), which creates complex expressions using JavaScript's type coercion; and JSFUCK (T10), which uses only six characters \texttt{[]()!+} to represent any JavaScript code, such as encoding the letter \texttt{"a"} as \texttt{(![]+[])[+!![]]}. 

\noindent {\bf Semantic-level Obfuscation.} These techniques fundamentally alter implementation logic while maintaining behavioral equivalence. String array techniques (T11, T12) collect all strings into arrays with complex decoding functions using XOR operations. JSON encoding (T13) transforms object literals into encoded strings that are decoded using XOR operations and parsed via \texttt{JSON.parse}. Regular expression encoding (T14) converts regex patterns into encoded character arrays that are reconstructed using \texttt{new RegExp}. Eval-based transformations (T15) introduce dynamic code generation through packed and encoded scripts. Control flow flattening (T16) converts linear code into switch-case state machines that execute in seemingly random order. Dead code insertion (T17) adds non-functional code branches and unreachable statements to increase complexity. 

\noindent {\bf Multi-layer Obfuscation.} These techniques combine multiple obfuscation strategies from different levels to create highly complex transformations. OB obfuscation (T18) implements commercial-grade protection by combining string arrays, control flow manipulation, and anti-debugging measures. LLM-based obfuscation (T19) leverages artificial intelligence to apply multiple techniques intelligently, creating code that appears naturally written but is heavily obfuscated.

\subsection{JavaScript Obfuscation Tools}
Modern JavaScript obfuscation tools implement diverse techniques with varying capabilities. We analyzed twelve representative obfuscation tools: MeSpl (Metasploit Framework JavaScript Obfuscator)~\cite{metasploit_js_obfuscation}, JSha (JShaman)~\cite{jshaman}, JSC (JScrambler)~\cite{jscrambler}, UGL (UglifyJS)~\cite{uglifyjs}, OBfu (Obfuscator.io)~\cite{ob}, JSV7 (jsjiami.com.v7)~\cite{jsjiami}, DAFT (DAFT logic Obfuscator)~\cite{daft}, TER (Terser)~\cite{terser}, FRE (FreejsObfuscator)~\cite{free}, STU (Stunnix JavaScript Obfuscator)~\cite{stunnix}, GNI (Gnirts Obfuscator)~\cite{gnirts}, and BEA (BeautifyTools Obfuscator)~\cite{beautify}. Table~\ref{tab:compare_obfuscation_tools} in Appendix~\ref{appendix:ob_cab} presents a comprehensive comparison of these tools' capabilities across all 20 obfuscation techniques.
The analysis reveals significant variation in implementation capabilities across different tools. Commercial tools like JSha achieve the highest technique implementation rates, supporting 16 out of 20 obfuscation categories (80\%), but fail to handle common techniques in real-world samples like AAEncode (T8), JJEncode (T9), JSFUCK (T10) and OB obfuscation (T18). However, JSha requires precise configuration tuning to achieve optimal results, creating usability barriers for practitioners. Mid-tier tools such as OBfu and JSC each implement 11 techniques (55\%), while specialized tools like MeSpl focus on specific use cases with more limited scope (6 techniques, 30\%). 

This comprehensive tool analysis reveals that modern obfuscation capabilities pose substantial challenges for security analysis. The widespread availability of powerful obfuscation methods enables attackers to easily access sophisticated techniques, while the diversity of available approaches makes it difficult to develop comprehensive countermeasures. This technological asymmetry between obfuscation and deobfuscation capabilities motivates the need for more advanced reverse engineering approaches.
\label{back}

\section{problem definition}
\subsection{A Real-World Motivating Example}
\label{subsec:motivation}
The complexity of modern JavaScript obfuscation techniques create substantial challenges for security analysis that existing deobfuscation tools cannot adequately address. 
To illustrate these challenges, we show a complex obfuscated wild script (Listing~\ref{lst:jsfiretruck_multilayer}) from the JSFireTruck campaign~\cite{unit42_jsfiretruck}, which infected 269,552 web pages in just one month.

JSFireTruck represents a particularly sophisticated obfuscation approach that exploits JavaScript's type coercion system to achieve extreme code transformation. It employs only six ASCII characters (\texttt{[]()!+}) by leveraging JavaScript's type coercion rules. For example, \texttt{+[]} converts empty array to 0, while \texttt{!![]} converts to boolean \texttt{True} then to number 1. The campaign combines this with \texttt{String.fromCharCode} functions and character array lookups, creating multi-layer obfuscation that decodes character arrays, processes JSFireTruck-encoded strings, and executes malicious code.

This sample demonstrates three critical challenges that existing tools cannot address comprehensively.

\noindent {\bf Challenge 1: Input Format Diversity.}
Existing deobfuscation tools show significant limitations in handling the diverse input formats encountered in real-world malicious samples. Tools like JStillery~\cite{jstillery} and IlluminateJS~\cite{illuminatejs} crash when encountering malformed syntax, mixed character encodings, or bundler-wrapped code common in real-world samples. These tools lack robust input validation and normalization capabilities, failing before analysis even begins. This fundamental weakness severely limits their real-world applicability where robust input preprocessing is essential.

\begin{lstlisting}[style=JavaScript, caption={Snippet of multi-layer obfuscated code.}, label={lst:jsfiretruck_multilayer}]
var _0x4f2a = [/* large character array with encoded strings */];
var _0x1b3c = String.fromCharCode(105, 102, 114, 97, 109, 101...);
var _0x2d4e = String.fromCharCode(115, 114, 99...);
// 13K characters of obfuscated code follows
[][(![]+[])[+[]]+([![]]+[][[]])[+!+[]+[+[]]]+(![]+[])[!+[]+!+[]]+(!![]+[])[+[]]+(!![]+[])[!+[]+!+[]+!+[]]+(!![]+[])
...  // truncated for brevity
\end{lstlisting}

\noindent {\bf Challenge 2: Multi-Layer Obfuscation Complexity.} The JSFireTruck example demonstrates the necessity of combining multiple analysis approaches to achieve comprehensive deobfuscation. The sample combines static patterns (character arrays, \texttt{String.fromCharCode} calls) with dynamic type coercion chains that require runtime evaluation. Static-only tools cannot resolve expressions like \texttt{+!![]} → 1 through complex coercion rules, while dynamic-only approaches struggle with the massive scale (269,552 samples) and security risks of executing malicious code. Furthermore, specialized tools like unjsfuck~\cite{unjsfuck} handle only the JSFUCK layer, leaving intermediate output like \texttt{\$[0] + \$[1].match(\$[2]|\$[3])} that still requires further process.

\noindent {\bf Challenge 3: Human Analysis Integration Gap.}
Even when tools successfully decode the obfuscation layers, the output commonly contains cryptic identifiers (\texttt{\_0x4f2a}, \texttt{\_0x1b3c}) and inconsistent formatting that impedes human analysis. Analysts need semantically meaningful variable names and professional formatting to quickly understand malicious behavior, but existing tools focus solely on functional correctness, ignoring human readability requirements essential for threat intelligence workflows.

\subsection{Problem Scope}
Our research focuses on ``narrow-sense obfuscation" - code transformations that make JavaScript difficult to understand while preserving its original functionality. This includes techniques like identifier renaming, string encoding, control flow flattening, and syntactic transformations that obscure code meaning but keep the code fully executable.

We exclude cryptographic protection schemes that require external decryption keys, as these represent fundamentally different security challenges beyond the scope of traditional deobfuscation. Instead, our approach targets the most common obfuscation techniques used in real-world malware campaigns like JSFireTruck, where the obfuscated code can be executed directly in any JavaScript environment without requiring additional secrets or external resources.
\label{problem}

\section{Design}
\OurTool adopts a modular approach that systematically combines static AST manipulation, dynamic execution monitoring, and LLM-enhanced processing. The modular design enables robust input handling, comprehensive obfuscation resolution, and human-readable output generation through a unified pipeline, delivering superior deobfuscation capabilities for real-world malware analysis.

\subsection{Overview}
As Figure~\ref{fig:system_overview_full} shows, \OurTool consists of a three-stage pipeline architecture: Preprocessor (\ding{192}), Deobfuscator (\ding{193}), and Humanizer (\ding{194}). Each module tackles a specific aspect of the deobfuscation challenge, collectively enabling effective processing of heavily obfuscated code. This modular design ensures robust handling of complex, multi-technique obfuscation while remaining flexible for future enhancements. The details of each module are described below.
\begin{figure*}[t]
\centering
\setcounter{figure}{0}  
\includegraphics[width=1\textwidth, keepaspectratio]{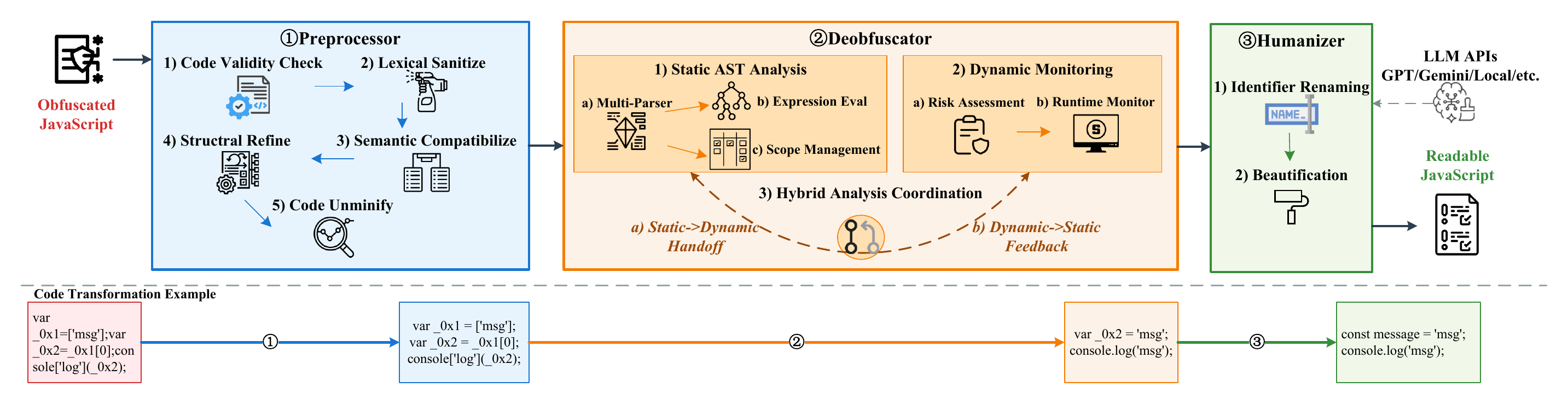}
\caption{Overview of \OurTool.}
\label{fig:system_overview_full}
\end{figure*}

    \noindent {\bf Preprocessor (\ding{192})}: The preprocessor validates input code integrity and normalizes syntax through lexical, semantic, and structural transformations. This ensures that subsequent analysis stages receive well-formed, processable code regardless of input variations. As shown in the code transformation example in Figure~\ref{fig:system_overview_full}, it handles malformed syntax and prepares the code structure for analysis.
    
    \noindent {\bf Deobfuscator (\ding{193})}: The deobfuscator applies targeted deobfuscation techniques using either static AST pattern matching for well-understood obfuscation patterns or dynamic execution tracing for complex, runtime-dependent transformations. The transformation example in Figure~\ref{fig:system_overview_full} demonstrates how obfuscated constructs are systematically resolved: array-based string encoding (\texttt{\_0x1 = ['msg']; \_0x2 = \_0x1[0]}) is simplified through static analysis to direct string assignment (\texttt{\_0x2 = 'msg'}).
    
    \noindent {\bf Humanizer (\ding{194})}: The humanizer employs LLMs to rename identifiers and improve semantic clarity, transforming mechanically deobfuscated code into human-readable form that facilitates further analysis.  As illustrated in the final output in Figure~\ref{fig:system_overview_full}, cryptic variable names are replaced with meaningful identifiers. Additionally, industry-standard formatting ensures a consistent professional code presentation that enhances readability and maintains compatibility with development environments.

\subsection{Preprocessor}
The preprocessor module serves as the critical foundation for reliable deobfuscation by ensuring that input code can be properly parsed and analyzed regardless of its initial state. Obfuscated JavaScript often contains malformed syntax, mixed encodings, and compatibility issues that can cause downstream analysis to fail. The preprocessor addresses these fundamental challenges through a systematic normalization pipeline.

\subsubsection{Code Validity Check}
We use Meriyah~\cite{meriyah} to handle malformed obfuscated JavaScript thanks to its strong fault tolerance and broad syntax support. Unlike traditional parsers that fail on broken or incomplete code, Meriyah generates complete ASTs even with syntax errors or problematic constructs. This capability proves indispensable for processing the diverse range of obfuscated inputs encountered in real-world scenarios where standard parsers would fail entirely.

\subsubsection{Lexical Sanitize}
We systematically normalize character encodings through a two-phase transformation process that resolves encoding conflicts in obfuscated JavaScript. 

The first phase converts legacy octal escape sequences to standardized hexadecimal format. This ensures parser compatibility across different JavaScript engines, since octal escapes are deprecated in strict mode and inconsistently supported~\cite{ecmascript2023}. We use pattern matching to identify three-digit octal patterns (e.g., \texttt{\textbackslash 302}), convert them to decimal values, and reformat them as hexadecimal representations (e.g., \texttt{\textbackslash xC2}).

The second phase addresses multi-byte UTF-8 character reconstruction. Non-ASCII characters require multiple bytes in UTF-8 encoding (e.g., Chinese characters need three bytes). Obfuscated code often splits these characters into separate \texttt{\textbackslash xHH} escape sequences that must be recombined for proper decoding. Our state machine parser identifies and groups consecutive escape sequences (e.g., \texttt{\textbackslash xE4\textbackslash xB8\textbackslash xAD}) up to four bytes. We validate hexadecimal pairs and attempt UTF-8 decoding to reconstruct the original characters into standardized Unicode escapes (e.g., \texttt{\textbackslash u4E2D}). When UTF-8 reconstruction fails, the system preserves individual hexadecimal escapes to prevent data corruption. Additionally, we extract JavaScript content from HTML containers to ensure consistent input formatting across different JavaScript engines.

\subsubsection{Semantic Compatibilize}
We replace legacy JavaScript constructs with cross-platform equivalents to ensure compatibility and strict-mode compliance. Legacy code often includes platform-specific directives and non-standard APIs that break modern parsers.

Our approach handles legacy conditional compilation through pattern recognition and standardization. We identify conditional compilation blocks (e.g., \texttt{/*@cc\_on...@*/}) and transform legacy platform detection methods into cross-platform equivalents. For example, legacy \texttt{@\_win32} conditions are converted to \texttt{navigator.platform} checks, and \texttt{@\_jscript} detection is replaced with feature detection using \texttt{'ActiveXObject' in window}. The transformation process also converts legacy conditional syntax (\texttt{@if}, \texttt{@else}, \texttt{@end}) into standard JavaScript control structures.

We address variable declaration inconsistencies through multi-phase processing. The system first identifies explicitly declared variables, then detects implicit global assignments that lack proper declarations, and automatically adds appropriate declaration keywords (\texttt{let}, \texttt{var}) to ensure strict-mode compatibility. This prevents runtime errors in modern JavaScript environments that enforce stricter scoping rules.

For problematic \texttt{delete} statements that violate strict-mode constraints, we implement selective removal based on context analysis. The system identifies \texttt{delete} operations on variables (rather than object properties), removes these invalid statements while preserving surrounding syntax structure, and cleans up residual punctuation to maintain code validity. This ensures that legacy code can execute properly in strict-mode environments without breaking existing program logic.

\subsubsection{Structural Refine}
We resolve declaration conflicts and code structure violations through AST scope chain traversal and systematic refactoring. Our implementation detects variable and function redeclarations by analyzing scope hierarchies, then refactors them into strict-mode compliant forms while preserving semantic meaning. This optimization improves code maintainability and prepares it for effective deobfuscation by eliminating redundant declarations and standardizing code structure.

\subsubsection{Code Unminify}
We integrate WebCrack~\cite{webcrack} to handle modern web applications that layer multiple obfuscation techniques on top of bundler-generated code. Our system automatically detects bundler-specific patterns (webpack~\cite{webpack}, browserify~\cite{browserify}) and applies specialized unpacking algorithms before general deobfuscation techniques. This preprocessing step is essential because attempting to deobfuscate webpack or browserify output without proper unpacking typically yields poor results due to nested obfuscation complexity.

\subsection{Deobfuscator}
We design a dual-strategy deobfuscation engine combining enhanced static AST analysis with controlled dynamic execution monitoring. Static analysis handles predictable obfuscation patterns, while dynamic analysis captures runtime-generated code through sandboxed execution. This hybrid approach provides comprehensive support across different obfuscation complexity levels.

\subsubsection{Static AST Analysis Deobfuscation}
Our static analysis approach introduces a comprehensive pattern recognition and transformation system that systematically handles diverse obfuscation techniques through enhanced AST manipulation. Unlike existing tools that focus on specific obfuscation patterns, our method provides unified handling of lexical, syntactic, and semantic-level obfuscations through enhanced expression evaluation, robust parser integration, and intelligent scope management. We implement these capabilities starting from JStillery's pattern recognition base:

\noindent {\bf a) Multi-Parser for Robust AST Construction.} We implement a fault-tolerant parsing system that integrates multiple JavaScript parsers (Esprima~\cite{esprima}, Acorn~\cite{acorn}, Babel~\cite{babel}) with automatic fallback mechanisms. When primary parsers encounter syntax errors or malformed code structures common in obfuscated JavaScript, the system automatically switches to alternative parsers with different error tolerance levels. This layered approach ensures successful AST construction even with heavily mangled input code.

\noindent {\bf b) Enhanced Expression Evaluation Engine.} We redesign the expression evaluation component to handle sophisticated obfuscation constructs.
Specifically, for \texttt{LogicalExpressions}, obfuscators frequently use nested \texttt{\&\&} and \texttt{||} chains to hide control flow logic. Our evaluation engine implements proper short-circuit evaluation that processes these operators left-to-right, stopping evaluation when the result is determined (e.g., \texttt{False \&\& anything} always returns \texttt{False}). This enables us to resolve complex boolean chains like \texttt{a \&\& b \&\& c || d} into their actual boolean values during static analysis.

Moreover, modern obfuscators exploit ES6 destructuring assignments to obscure variable bindings. For example, instead of simple assignments like \texttt{var a = 1; var b = 2;}, obfuscators generate complex patterns such as \texttt{[a, b, c] = [getValue(), obj.prop, func.call(this)]} or nested destructuring like \texttt{[x, [y, z]] = [arr[0], [obj.method(), computed()]]}. These patterns break traditional variable tracking because the assignment targets are not immediately visible. We extend \texttt{AssignmentExpression} processing to handle \texttt{ArrayPattern} destructuring by parsing the left-hand pattern structure, recursively traversing nested array patterns, mapping each element position to its corresponding right-hand value, and creating individual variable bindings for each destructured element while preserving the evaluation order. \looseness=-1

For \texttt{UnaryExpressions}, obfuscators often apply \texttt{typeof} or negation operators to global variables in conditional checks that depend on runtime environment detection. Common obfuscation patterns include \texttt{typeof window !== 'undefined' \&\& window.document} for browser detection, \texttt{!!(typeof global === 'object' \&\& global)} for Node.js environment checks, or \texttt{typeof navigator !== 'undefined' \&\& navigator.userAgent} for user agent sniffing. Static evaluation of these expressions would break environment-specific code paths. We implement an \texttt{excludedNames} whitelist containing critical globals (\texttt{window}, \texttt{document}, \texttt{navigator}, \texttt{global}, \texttt{process}) that should never be statically evaluated. When processing expressions like \texttt{typeof window} or \texttt{!document}, the system preserves these expressions unchanged to maintain correct runtime behavior across different JavaScript environments.

\noindent {\bf c) Intelligent Scope and Dependency Management.} Obfuscated code breaks variable resolution through indirect function calls and complex reference chains that standard scope analysis cannot follow. We address this through two integrated mechanisms: dependency tracking for indirect function calls and enhanced scope chain management for variable resolution.

Our \texttt{functionneed} mechanism records dependency relationships when encountering calls where the callee is determined by variable lookup or property access (e.g., \texttt{obj[funcName](args)}), ensuring proper resolution before evaluation. Meanwhile, our scope traversal algorithm maintains a scope chain that tracks variable bindings across function boundaries, preserving parent scope access while handling variable shadowing to prevent incorrect substitution that could change program semantics.

\subsubsection{Dynamic Monitoring Execution Deobfuscation}
For obfuscation techniques that resist static analysis, we implement a controlled execution monitoring system that safely captures runtime behavior while maintaining security isolation, inspired by dynamic malware analysis approaches~\cite{kolbitsch2012rozzle}. Our approach operates through two integrated phases: pre-execution risk assessment and controlled runtime monitoring.

\noindent {\bf a) Pre-Execution Risk Assessment.} Before attempting dynamic execution, we analyze code patterns to identify potential execution hazards. The safety guard system scans for dangerous keyword combinations (\texttt{push}, \texttt{shift}, \texttt{eval}, \texttt{await}) that indicate complex nested obfuscation schemes prone to infinite loops or recursive execution deadlocks. This pattern-based risk assessment determines whether code segments can be safely executed or require alternative handling strategies.

We enhance risk assessment by mapping function dependencies to trace calls between obfuscated functions. This identifies critical functions needing dynamic resolution and their relationships, enabling understanding of execution flows and prioritizing order to avoid circular dependencies.

\noindent {\bf b) Controlled Runtime Monitoring.} We use Node.js \texttt{vm.runInNewContext} to create isolated execution contexts. Each obfuscated code segment runs in a separate sandboxed VM instance with no access to the file system, network, or global objects. The environment exposes only essential built-in objects while blocking dangerous APIs like \texttt{require} and \texttt{process}.

We execute obfuscated code segments in the isolated environment and capture their final execution results. The sandbox execution allows us to safely evaluate expressions that cannot be statically analyzed, such as dynamically constructed strings or runtime-dependent calculations. For example, when processing string array deobfuscation, we execute the array construction and access operations to obtain the actual string values. We implement comprehensive security mechanisms including execution timeouts to prevent hanging processes, recursion depth limits to prevent infinite loops, and memory monitoring to prevent resource exhaustion attacks. When runtime execution encounters unresolved references, the system consults static scope analysis results to inject appropriate variable definitions. Our system primarily executes limited-scope code patterns (pure function calls, IIFE executions, object method calls) to minimize exposure to malicious operations, though we recommend additional isolation for production environments.

\subsubsection{Hybrid Analysis Coordination} We implement coordination mechanisms to integrate static and dynamic analysis results through bidirectional information flow.

\noindent {\bf a) Static-to-Dynamic Handoff.} When static analysis encounters expressions that cannot be safely evaluated, \OurTool's \texttt{canbetransformed} marking mechanism identifies these expressions and packages their context information for dynamic execution. \texttt{CallExpression} requiring dynamic resolution typically involves: (1) function calls where the callee is determined by variable lookup or property access (e.g., \texttt{objfuncName}), (2) calls to functions that perform runtime code generation or string manipulation that affects program structure, and (3) calls that depend on runtime state or external context not available during static analysis. The system transfers the \texttt{CallExpression} context, scope information, and dependency relationships to the dynamic execution monitor, providing necessary context for safe runtime analysis. \looseness=-1

\noindent {\bf b) Dynamic-to-Static Feedback Integration.} Dynamic execution results are validated and integrated back into the static AST through type-aware processing that maintains semantic correctness. Simple data types (strings, numbers, booleans) are directly converted to literal AST nodes, while function results are parsed back into proper \texttt{FunctionExpression} nodes. Complex objects undergo JSON serialization before integration to ensure safe representation in the AST structure.

The system maintains AST structural integrity through careful node replacement that preserves surrounding syntactic context and control flow relationships. Additionally, the system updates variable bindings in the scope chain, replaces complex expressions with their resolved literal values, and triggers re-analysis of dependent code sections that may now be statically evaluable with the new information.

\subsection{Humanizer}
While the Deobfuscator successfully restores program logic and control flow, the resulting code often retains cryptic identifiers and inconsistent formatting that impede human analysis. To address this limitation, we implement a post-processing pipeline that transforms mechanically correct but unreadable code into professionally formatted, semantically clear representations suitable for security analysis and reverse engineering workflows.

\subsubsection{Identifier Renaming}
We integrate LLM capabilities for intelligent variable and function renaming that replaces meaningless obfuscated identifiers with semantically meaningful names. Our implementation leverages the core AST-based renaming approach from HumanifyJS~\cite{humanify}, adapting it with multi-provider LLM support and optimized context management for deobfuscation workflows.

\subsubsection{Beautification}
We integrate Prettier~\cite{prettier} formatting capabilities to ensure deobfuscated output meets professional development standards. The formatting pipeline applies consistent indentation schemes, standardized bracket placement, normalized quote usage, and appropriate semicolon handling according to established JavaScript coding conventions. This transformation ensures that the final output integrates seamlessly with development workflows and maintains compatibility with standard code analysis tools used in security research environments.

\label{design}

\section{Evaluation}
Our evaluation encompasses eight complementary assessment dimensions to comprehensively measure deobfuscation performance, covering deobfuscation capability, result correctness, semantic consistency, code simplification, ablation study, large-scale effectiveness and performance, LLM-based realiability and readability validation and user study.

\subsection{Dataset}
Our evaluation employs three complementary datasets to comprehensively assess deobfuscation performance across different threat scenarios: two datasets (MalJS, BenignJS) curated by us and one dataset (CombiBench) from JsDeobsBench~\cite{jsdeobsbench2025}. 
Table~\ref{tab:datasets} provides an overview of the datasets.

\begin{table}[htbp]
  \centering
  \caption{Evaluation datasets overview.}
  \label{tab:datasets}
  \begin{tabular}{lrrp{2.4cm}}
  \toprule
    \textbf{Dataset} & \textbf{\# Sample} & \textbf{Avg. Size} & \textbf{Source} \\
    \midrule
    \rowcolor[HTML]{EFEFEF} \textbf{MalJS} & 23,212 & 391.78 KB & A corporate partner \\
    \textbf{BenignJS} & 21,209 & 41.40 KB & GitHub + Web \\
    \rowcolor[HTML]{EFEFEF} \textbf{CombiBench} & 1,296 & 8.06 KB & JsDeobsBench \\
  \bottomrule
  \end{tabular}
\end{table}
\subsubsection{Obfuscation Scoring and Sample Selection.}
To ensure dataset's quality, we implemented a rigorous preprocessing pipeline.
Inspired by previous work~\cite{chai2022invoke}, we applied obfuscation scoring methodology to quantify obfuscation levels in our dataset. Following the four-category classification in Section~\ref{sec:js_obfuscation}, we assign scores of 1, 2, 3, and 4 points to lexical-level, syntactic-level, semantic-level, and multi-layer obfuscation techniques respectively, with each technique type contributing only once per script.

\subsubsection{Malicious JavaScript Dataset (MalJS)}
We collected 10,664,628 wild malicious JavaScript samples from July 3, 2019 to July 9, 2025 with the help of a leading cybersecurity company. 
We first utilized the Meriyah JavaScript parser for syntax validation, filtering out corrupted files and non-JavaScript content. This step retained 8,451,998 syntax-valid samples. To address highly similar samples within malware families, we performed content-based deduplication by normalizing scripts (replacing all string literals with placeholders, removing whitespace and newlines) and removing samples with identical normalized content, resulting in 4,470,565 unique samples. The dataset shows diverse obfuscation complexity with samples using 0-16 techniques (average 8.32). For representative evaluation, we selected samples across the full complexity spectrum (scores 0-35): 700 samples each for scores 1-33, and all available samples for the highest complexity levels (3 and 109 samples for score 34 and 35 respectively), totaling 23,212 samples.


\subsubsection{Benign JavaScript Dataset (BenignJS)}
This dataset contains 21,209 benign samples from two sources. First, we retained 19,209 samples sampled from the Top 1K popular websites from Tranco~\cite{tranco2024} with obfuscation scores ≥ 1.
Second, we curated 2,000 clean samples verified to be unobfuscated. These include 1,000 samples from popular open-source JavaScript projects on GitHub (trending repositories with over 3,000 stars) collected between April 15 and May 1, 2025, and 1,000 from PublicWWW~\cite{publicwww} crawling of legitimate websites. All clean samples were manually checked for absence of obfuscation. For evaluation in Sections~\ref{sec:result_correct} and \ref{sec:semantic_consistency}, we apply systematic obfuscation to these clean samples across four categories (500 samples each). Each sample is obfuscated using randomly selected techniques from its category, with at least one and up to all available techniques applied, then processed through \OurTool's three-stage pipeline. 


\subsubsection{Benchmark Evaluation Dataset (CombiBench)}
To ensure a fair comparison with existing deobfuscation methods, we also evaluate our approach on the JsDeObsBench benchmark, which is an evaluation framework specifically designed for JavaScript deobfuscation research. 
We use the complex combination obfuscation subset, termed \textit{CombiBench}, containing 1,296 samples that combine multiple obfuscation techniques from different categories to represent the most challenging evaluation scenarios. 

\subsection{Experimental Setup}

Complete experimental configurations and software dependencies are detailed in Appendix~\ref{appendix:artifact}. 
We employed Joern v1.1.1548 for code property graph generation with Neo4j 4.4.12 (8 GB heap allocation). Additional software included custom JavaScript parsers built on Esprima v4.0.1 for AST manipulation.

For deobfuscated capability evaluation, we applied each of the 20 obfuscation techniques from our categorization (T0-T19) to the test code, generating 20 distinct obfuscated versions. These obfuscated samples were then processed by:
\begin{itemize}
\item \textit{Ten state-of-the-art deobfuscation tools}: JSNice (JSN)~\cite{raychev2015predicting}, Dev-coco (DEV)~\cite{devcoco2024jsdeobfuscator}, JStillery (JST)~\cite{jstillery}, illuminateJS (ILL)~\cite{illuminatejs}, JS-Deob (JSD)~\cite{kuizuo2024jsdeobfuscator}, JSbeautifier.org (JSB)~\cite{lielmanis2024jsbeautify}, synchrony (SYN)~\cite{synchrony2024}, JSDE (JSDec)~\cite{hax0r31337_jsdec}, JS-deobfuscator (SDS)~\cite{bensb_deobfuscator_io}, and JSIMPO (JSI)~\cite{jsimpo}
\item \textit{Three LLM-based solutions}: GPT-o1 (GPT), Gemini-2.0-Flash (GEM), and DeepSeek-R1 (DSR)
\item \textit{Our tool}: \OurTool
\end{itemize}

For LLM integration, we utilized multiple state-of-the-art models via a unified API service: GPT-o3, DeepSeek-R1, Gemini 2.5 Pro, and Claude 3.7 Sonnet. Model-specific configurations included temperature settings of 0.7 for GPT-o3 and 0.3 for other models, with maximum token limits set to 4,000 across all models. API requests were implemented using asynchronous processing with aiohttp~\cite{aiohttp2025} for efficiency, incorporating exponential backoff retry mechanisms and proper error handling to ensure robust and reproducible results. 



\subsection{Deobfuscation Capability}
\label{sec:deob_capability}
To evaluate the deobfuscation capabilities of existing tools, we conducted extensive experiments using the categorized obfuscation techniques. We selected a representative JavaScript code snippet (in Appendix~\ref{appendix:obfuscator_options}) and applied all 20 obfuscation techniques individually to generate test cases. This code snippet encompasses various JavaScript constructs including function declarations, variable assignments, string operations, arithmetic expressions, boolean logic, and conditional statements, making it suitable for testing different obfuscation techniques. Additionally, we developed comprehensive demonstration examples for each obfuscation technique to ensure thorough evaluation.
The prompt used for LLM-based deobfuscation is provided in Appendix~\ref{appendix:llm_prompt_deobfuscation}.

\noindent {\bf Results.} 
Table~\ref{tab:compare_deobfuscation_tools} presents the key comparison results for all evaluated tools. Our analysis reveals significant variations in deobfuscation capabilities across different tool categories and obfuscation techniques, with multi-layered techniques like LLM obfuscation and commercial OB tools posing the greatest challenges for existing deobfuscation solutions.

\begin{table*}[htbp]
  \centering
  \begin{threeparttable}
  \caption{Comprehensive comparison of deobfuscation capability of different tools.}
  \label{tab:compare_deobfuscation_tools}
  \begin{tabular}{p{1.5cm}p{2cm}p{0.5cm}p{0.5cm}p{0.5cm}p{0.5cm}p{0.5cm}p{0.5cm}p{0.5cm}p{0.5cm}p{0.5cm}p{0.5cm}p{0.5cm}p{0.5cm}p{0.5cm}p{0.5cm}}
  \toprule
    \textbf{Type} & \textbf{Method} & \textbf{JSN}& \textbf{DEV}& \textbf{JST}& \textbf{ILL}& \textbf{JSD}& \textbf{JSB}& \textbf{SYN}& \textbf{JSDec}& \textbf{SDS}& \textbf{JSI}& \textbf{GEM}& \textbf{GPT}& \textbf{DSR}&  \textbf{Ours}\\
    \hline
    \multirow{5}{*}{\centering Lexical} & T0:Rename &  &  &  &  &  &  &  &  &  &  &$\checkmark$ & $\checkmark$ & $\checkmark$ & $\checkmark$\\
    \rowcolor[HTML]{EFEFEF} \cellcolor{white}                       & T1:Indirect & $\checkmark$ & $\checkmark$ & $\checkmark$ & $\checkmark$ & $\checkmark$ &  & $\checkmark$ & $\checkmark$ & $\checkmark$ &  & $\checkmark$ & $\checkmark$ & $\checkmark$ & $\checkmark$\\
                            & T2:Arithmetize &  & $\checkmark$ & $\checkmark$ & $\checkmark$ &  &  &  &  &  &  & $\checkmark$ & $\checkmark$ & $\circ$ & $\checkmark$\\
    \rowcolor[HTML]{EFEFEF} \cellcolor{white}                       & T3:StringEncode & $\checkmark$ & $\checkmark$ & $\checkmark$ & $\checkmark$ & $\checkmark$ & $\checkmark$ & $\checkmark$ & $\checkmark$ & $\checkmark$ &  & $\checkmark$ & $\checkmark$ & $\checkmark$ & $\checkmark$\\
                            & T4:BooleanEncode &  & $\checkmark$ & $\checkmark$ & $\checkmark$ & $\checkmark$ &  & $\checkmark$ & $\checkmark$ & $\checkmark$ &  & $\checkmark$ & $\checkmark$ & $\checkmark$ & $\checkmark$\\
    \hline
    \rowcolor[HTML]{EFEFEF} \cellcolor{white} \multirow{6}{*}[1.5em]{\centering Syntactic}    & T5:Expr2Function &  &  & $\checkmark$ & $\checkmark$ &  &  &  &  &  & $\circ$ & $\checkmark$ & $\checkmark$ & $\checkmark$ & $\checkmark$\\
                            & T6:Assign2Function &  &  & $\checkmark$ & $\checkmark$ &  &  &  & $\checkmark$ & $\checkmark$ & $\circ$ & $\checkmark$ & $\checkmark$ & $\checkmark$ &  $\checkmark$\\
    \rowcolor[HTML]{EFEFEF}  \cellcolor{white}                    & T7:Reverse &  &  &  & $\checkmark$ &  &  &  &  & $\checkmark$ &  & $\checkmark$ & $\checkmark$ & $\checkmark$ & $\checkmark$\\
                            & T8:AAEncode &  &  & $\checkmark$ & $\checkmark$ &  &  &  &  &  &   & $\circ$ & $\circ$ &  & $\checkmark$\\
    \rowcolor[HTML]{EFEFEF} \cellcolor{white}                       & T9:JJEncode &  &  &  &  &  &  &  &  &  &   & $\circ$ &  & $\circ$ & $\checkmark$\\
                            & T10:JSFUCK &  &  &  &  &  &  &  &  &  &  & $\circ$ &  & $\circ$ & $\checkmark$\\
    \hline
    \rowcolor[HTML]{EFEFEF} \cellcolor{white}\multirow{7}{*}{\centering Semantic}  & T11:Arrayize & $\circ$ & $\circ$ &  & $\checkmark$ &  & $\circ$ &  &  & $\checkmark$ &  &  & $\checkmark$ & $\checkmark$ & $\checkmark$\\
                            & T12:StrArrEncode &  & $\checkmark$ & $\circ$ & $\checkmark$ &  &  &  &  & $\checkmark$ & $\circ$ &  & $\checkmark$ & $\checkmark$ & $\checkmark$\\
    \rowcolor[HTML]{EFEFEF} \cellcolor{white}                       & T13:JSONEncode &  &  & $\circ$ &  &  &  &  &  &  & $\circ$ &  & $\checkmark$ & $\circ$ & $\checkmark$\\
                            & T14:RegexpEncode &  &  & $\circ$ &  &  &  &  &  &  & $\circ$ &  & $\checkmark$ & $\checkmark$ & $\checkmark$\\
    \rowcolor[HTML]{EFEFEF} \cellcolor{white}                       & T15:Eval &  & $\checkmark$ & $\circ$ & $\circ$ &  &  &  & $\checkmark$ &  & $\circ$ & $\checkmark$ & $\checkmark$ & $\checkmark$ & $\checkmark$\\
                            & T16:Flattern & $\circ$ & $\checkmark$ & $\circ$ & $\circ$ & $\checkmark$ &  &  &  &  & $\checkmark$ & $\checkmark$ & $\checkmark$ & $\checkmark$ &  $\checkmark$\\
    \rowcolor[HTML]{EFEFEF} \cellcolor{white}                       & T17:Insert & $\circ$ &  & $\circ$ &  &  &  &  & $\checkmark$ & $\circ$ &  &  &  &  & $\checkmark$\\
    \hline
    \multirow{2}{*}[2pt]{Multi-Layered} & T18:OB &  &  &  &  &  &  &  & $\circ$ & $\circ$ & $\checkmark$ & $\circ$ & $\circ$ & $\circ$ & $\checkmark$\\
    \rowcolor[HTML]{EFEFEF} \cellcolor{white}                        & T19:LLM & $\circ$ & $\circ$ & $\checkmark$ & $\checkmark$ & $\circ$ & $\circ$ & $\circ$ & $\checkmark$ &  &  & $\checkmark$ & $\checkmark$ & $\checkmark$ & $\checkmark$\\
    \bottomrule
  \end{tabular}
    \begin{tablenotes}
    \footnotesize
    \item Note: $\checkmark$ = fully successful, $\circ$ = partially successful, blank = fully unsupported.
    \end{tablenotes}
\end{threeparttable}
\end{table*}

Traditional tools demonstrate fundamental limitations across obfuscation categories. Tools like JSN and DEV struggle with lexical transformations(T0) and fail on advanced syntactic transformations like expression-to-function conversion. More sophisticated tools (JST, ILL) handle syntactic techniques moderately but face challenges with semantic-level obfuscation. JSI shows specialized but limited capabilities, reliably handling control-flow flattening (T16) and OB obfuscation (T18) but failing on most other techniques due to strict pattern matching requirements. Exotic encoding techniques (T8, T9, T10) and advanced semantic transformations like JSON encoding (T13) challenge traditional tools due to limited semantic analysis capabilities. Insertion technique (T17) and OB obfuscation (T18) represent the most difficult categories, with most existing tools showing complete failure.

LLM-based solutions show superior performance compared to traditional tools. GPT excels most lexical and syntactic techniques while showing strong capabilities in semantic transformations. However, even these advanced approaches struggle with the most sophisticated obfuscation methods. 

In contrast, \OurTool achieves success across all 20 obfuscation techniques, including the most challenging T17 (insertion) and T18 (OB) techniques. This comprehensive evaluation highlights the significant limitations of existing deobfuscation approaches and demonstrates the need for solutions capable of handling modern obfuscation complexity.

\subsection{Result Correctness}
\label{sec:result_correct}
We evaluate result correctness by assessing whether deobfuscated code maintains syntactic validity with the original code. We conduct automated parser validation on all BenignJS and MalJS samples, and manual verification on 100 BenignJS and 300 MalJS samples for real-world effectiveness validation.
Based on their deobfuscation capability (Section~\ref{sec:deob_capability}), we selected six best-performing methods for comparison: JST, ILL, SDS, GPT, GEM, and DSR.

Manual verification involves: (1) reviewing parser error messages to categorize failure types (e.g., missing brackets, invalid identifiers); (2) examining deobfuscated code structure to identify syntactic malformations that may not trigger parser errors; (3) validating ECMAScript compliance. Tools that fail to process samples or simply output the input content with minor formatting changes are considered unsuccessful. Samples pass only when both automated parser validation and manual verification succeed.

\noindent {\bf Results.}
Table~\ref{tab:Result_Correctness_Semantic_Consistency} presents the correctness evaluation results across 400 samples (100 BenignJS + 300 MalJS) for detailed tool comparison. \OurTool achieved 100\% syntactic correctness across both BenignJS and MalJS datasets, with zero parsing errors confirmed through manual verification.
Traditional tools showed different results: SDS processed 304 samples with full correctness, while JST and ILL processed fewer samples (220 and 164) but maintained high accuracy. LLM-based methods had mixed performance: GEM processed 396 samples but only 180 were correct, while GPT and DSR processed 212 and 160 samples respectively.

Manual verification of syntactically valid outputs confirmed \OurTool's 100\% result correctness across both datasets. Traditional tools maintained high accuracy on successfully processed samples: SDS achieved 100\% (76/76), ILL maintained 100\% (41/41), and JST reached 85.45\% (47/55). LLM-based approaches showed lower performance: DSR obtained 87.50\% (35/40), GPT achieved 66.04\% (35/53), and GEM reached 45.45\% (45/99).


\begin{table}[htbp]
  \centering
  \caption{Correctness and consistency evaluation results.}
  \label{tab:Result_Correctness_Semantic_Consistency}
  \begin{tabular}{l*{7}{p{0.5cm}}}
  \toprule
    \textbf{Metrics} & \textbf{JST} & \textbf{ILL} & \textbf{SDS} & \textbf{GPT} & \textbf{GEM} & \textbf{DSR} & \textbf{Ours} \\
    \midrule
     \rowcolor[HTML]{EFEFEF} \#Deob (/400) & 220 & 164 & 304 & 212 & 396 & 160 & \textbf{400} \\
    \#Correct(/400) & 188 & 164 & 304 & 140 & 180 & 140 & \textbf{400} \\
     \rowcolor[HTML]{EFEFEF} \#Consist (/100) & 26 & 20 & 40 & 26 & 25 & 19 & \textbf{83} \\
    CFG Sim. (\%) & 58.06 & 74.95 & 67.04 & 58.35 & 76.75 & 92.75 & \textbf{93.78} \\
     \rowcolor[HTML]{EFEFEF} DDG Pres. (\%) & 59.00 & 79.20 & 63.52 & 55.50 & 78.50 & 89.50 & \textbf{95.84} \\
    \bottomrule
  \end{tabular}
\end{table}


\subsection{Semantic Consistency}
\label{sec:semantic_consistency}
We verify that deobfuscated code maintains identical runtime behavior and logical flow compared to original samples using automated graph-based analysis with manual verification.
Specifically, Joern generates CFGs and DDGs for structural comparison. The evaluation employs a two-tier approach: for graphs with $\leq$40 nodes, the Maximum Common Subgraph (MCS) algorithm computes similarity as the ratio of MCS size to maximum node count. For larger graphs ($>$40 nodes), fast approximation based on node/edge count differences avoids exponential complexity. When similarity scores fall below 75\%, we conduct manual analysis. Samples are marked consistent only when manual verification confirms preserved program behaviors.

\noindent {\bf Results.} Table~\ref{tab:Result_Correctness_Semantic_Consistency} presents the semantic consistency evaluation results. \OurTool achieved 83\% semantic consistency (83 out of 100 samples), outperforming GPT (74.29\%, 26/35 samples) and JST (55.32\%, 26/47 samples). Additionally, structural analysis shows \OurTool maintains superior code fidelity with 93.78\% CFG similarity and 95.84\% DDG preservation, compared to GPT's 58.35\% and 55.50\%. 

To verify this high consistency rate, we applied a manual verification process with specific criteria. For the 83 consistent samples, we confirmed: (1) Identical execution paths with conditional branches, loops, and function calls following the same logical sequence; (2) Preserved variable relationships including assignments, scoping, and data dependencies; (3) Equivalent API usage maintaining identical parameters and invocation order, allowing syntactic variations like \texttt{obj["prop"]} vs \texttt{obj.prop}; (4) Consistent computational results. Appendix~\ref{appendix:cfg_ddg} shows a representative consistent sample\footnote{MD5: 000c5123cd553d14cd49d22022fcc9c5} demonstrating preserved control flow and data dependencies.

\noindent {\bf Failure Analysis.}
To address the 17\% semantic inconsistency cases, we conducted failure analysis.
For the 17 inconsistent samples, our analysis revealed: (1) Altered control flow (8 samples): missing or additional conditional branches, modified loop conditions, or changed function call sequences; (2) Broken variable dependencies (6 samples): incorrect variable scoping, missing assignments, or spurious data relationships; (3) Modified API behavior (4 samples): changed function parameters, altered invocation order, or missing critical API calls; (4) Computational discrepancies (3 samples): different mathematical results, incorrect string operations, or modified logical outcomes. Some samples exhibited multiple inconsistency types, with 4 samples showing overlapping issues. The primary cause stems from LLM-based variable renaming introducing semantic errors. When the LLM performs automatic variable renaming, it may inadequately analyze dynamic variable assignments and scoping, leading to undefined variables that break the original call chains and functionality. A detailed failure case is provided in Appendix~\ref{appendix:failure_case}.

\subsection{Code Simplification}
\label{sec:code_simplification}
As in previous work~\cite{jsdeobsbench2025}, we adopt the Halstead length~\cite{halstead1977} to measure code complexity and the Halstead Length Reduction (HLR) score to evaluate the effectiveness of deobfuscation tools. 
The HLR score is calculated as $(HLoC_{obf} - HLoC_{deobf}) / HLoC_{obf}$, where higher values indicate greater complexity reduction. According to the definition, the HLR score ranges from 0 to 1, where higher values indicate greater complexity reduction. We also evaluate Halstead Effort Reduction (HER) score, which measures the reduction in mental effort required to understand code. 
The evaluation uses the CombiBench dataset, comparing against SDS and SYN as the primary baseline tools evaluated in the JsDeObsBench.

\noindent {\bf Results.}
Table~\ref{tab:hlr_comparison} presents the complexity reduction scores following the JsDeObsBench evaluation setup. We achieved an HLR score of 0.8820 and HER of 0.9291, demonstrating substantial complexity reduction capabilities. SDS~\cite{bensb_deobfuscator_io} achieved minimal simplification with HLR and HER score of only 0.0015 and 0.1174. While SYN~\cite{synchrony2024} showed better performance with HLR and HER score of 0.7889 and 0.8736, it can only handle specific obfuscation combinations and lacks generalization to diverse real-world obfuscation patterns. Our results demonstrate that \OurTool achieves substantial complexity reduction, confirming our method's effectiveness in transforming obfuscated code into obvious forms.

\begin{table}[htbp]
\centering
\caption{Results of HLR and HER score comparison.}
\label{tab:hlr_comparison}
\begin{tabular}{ccc}
\toprule
Tool & HLR Score & HER Score\\
\midrule
\OurTool & \textbf{0.8820} & \textbf{0.9291}\\
\rowcolor[HTML]{EFEFEF}synchrony (SYN) & 0.7889 & 0.8736\\
JS-deobfuscator (SDS) & 0.0015 & 0.1174\\
\bottomrule
\end{tabular}
\end{table}


\subsection{Ablation Study}
\label{sec:ablation}
To quantify each module's contribution, we evaluated four configurations on 1,000 randomly selected MalJS samples (1KB to 5.49MB, average 82.23KB): (1) Preprocessor only, (2) Preprocessor + Static Analysis, (3) Preprocessor + Deobfuscator (Static + Dynamic), and (4) Full system (with Humanizer).

\noindent {\bf Results.}
Table~\ref{tab:ablation} shows the ablation results. Processing times remain similar across most configurations (9.16-9.26 seconds) due to fixed-cost operations like file I/O and AST parsing. The incremental overhead from static and dynamic modules is relatively small since most samples do not trigger complex obfuscation-specific processing. The full system requires 87.31 seconds due to LLM API latency but achieves optimal deobfuscation quality.

Table~\ref{tab:ablation_detailed} presents detailed deobfuscation capability breakdown across all 20 obfuscation techniques. The Preprocessor handles 0/20 techniques (syntactic normalization only). Static analysis enables 14/20 techniques. The combined Preprocessor + Deobfuscator (static + dynamic) handles 19/20 techniques, missing only identifier renaming. The full system achieves complete 20/20 coverage.

For code simplification, we measured Halstead Length Reduction (HLR) on real-world samples. Static analysis achieves 3.52\% HLR. The combined static+dynamic approach achieves 3.40\% HLR (slightly lower due to additional code introduced during dynamic analysis).  The full system achieves 5.66\% HLR, significantly lower than the 88.20\% achieved on CombiBench (Section~\ref{sec:code_simplification}) due to real-world complexity.
\begin{table}[htbp]
\centering
\caption{Ablation study showing module contributions.}
\label{tab:ablation}
\begin{tabular}{lccc}
\toprule
Configuration & Avg. Time (s) & Techniques & HLR (\%) \\
\midrule
Preprocessor & 9.26 & 0/20 & 0.00 \\
\rowcolor[HTML]{EFEFEF} + Static & 9.16 & 14/20 & 3.52 \\
+ Dynamic & 9.23 & 19/20 & 3.40 \\
\rowcolor[HTML]{EFEFEF} + LLM & 87.31 & \textbf{20/20} & \textbf{5.66} \\
\bottomrule
\end{tabular}
\end{table}

\begin{table}[htbp]
\centering
\begin{threeparttable}
\caption{Deobfuscation capability by module configuration.}
\label{tab:ablation_detailed}
\begin{tabular}{lp{1.4cm}p{1.4cm}p{1.4cm}p{1.4cm}}
\toprule
\textbf{Technique} & \textbf{Prep} & \textbf{+Static} & \textbf{+Dynamic} & \textbf{+LLM} \\
\midrule
\rowcolor[HTML]{EFEFEF} T0 &  &  &  & $\checkmark^*$ \\
 T1-T8 &  & $\checkmark$ & $\checkmark$ & $\checkmark$ \\
\rowcolor[HTML]{EFEFEF} T9-T10 &  &  & $\checkmark^*$ & $\checkmark$ \\
 T11-T14 &  & $\checkmark$ & $\checkmark$ & $\checkmark$ \\
\rowcolor[HTML]{EFEFEF} T15-T16 &  &  & $\checkmark^*$ & $\checkmark$ \\
 T17 &  & $\checkmark$ & $\checkmark$ & $\checkmark$ \\
\rowcolor[HTML]{EFEFEF} T18 &  &  & $\checkmark^*$ & $\checkmark$ \\
 T19 &  & $\checkmark$ & $\checkmark$ & $\checkmark$ \\
\midrule
\textbf{Total} & \textbf{0/20} & \textbf{14/20} & \textbf{19/20} & \textbf{20/20} \\
\bottomrule
\end{tabular}
  \begin{tablenotes}
    \footnotesize
    \item [*] indicates newly enabled capability in this configuration.
    \end{tablenotes}
    \end{threeparttable}
\end{table}

\subsection{Large-Scale Effectiveness and Performance}
To validate practical code simplification effectiveness and processing efficiency, we conducted large-scale evaluation on MalJS (23,212 samples) and BenignJS (21,209 samples) datasets. We performed full-scale entropy analysis on both datasets to assess effectiveness, and comprehensive performance evaluation on the complete MalJS dataset.

We employed two complementary entropy metrics to quantify deobfuscation quality: Code Text Entropy measures textual randomness through character and word frequency analysis, while AST Entropy evaluates structural complexity using node types, connectivity, depth, and edge relationships. We selected JST and ILL as comparison baselines due to their deobfuscation capabilities and batch processing support. For fair comparison, we measured only \OurTool's Preprocessor and Deobfuscator, excluding the Humanizer. Detailed formulations are provided in Appendix~\ref{appendix:entropy_fomulation}.

\begin{figure}[htbp]
\includegraphics[width=0.49\textwidth, keepaspectratio]{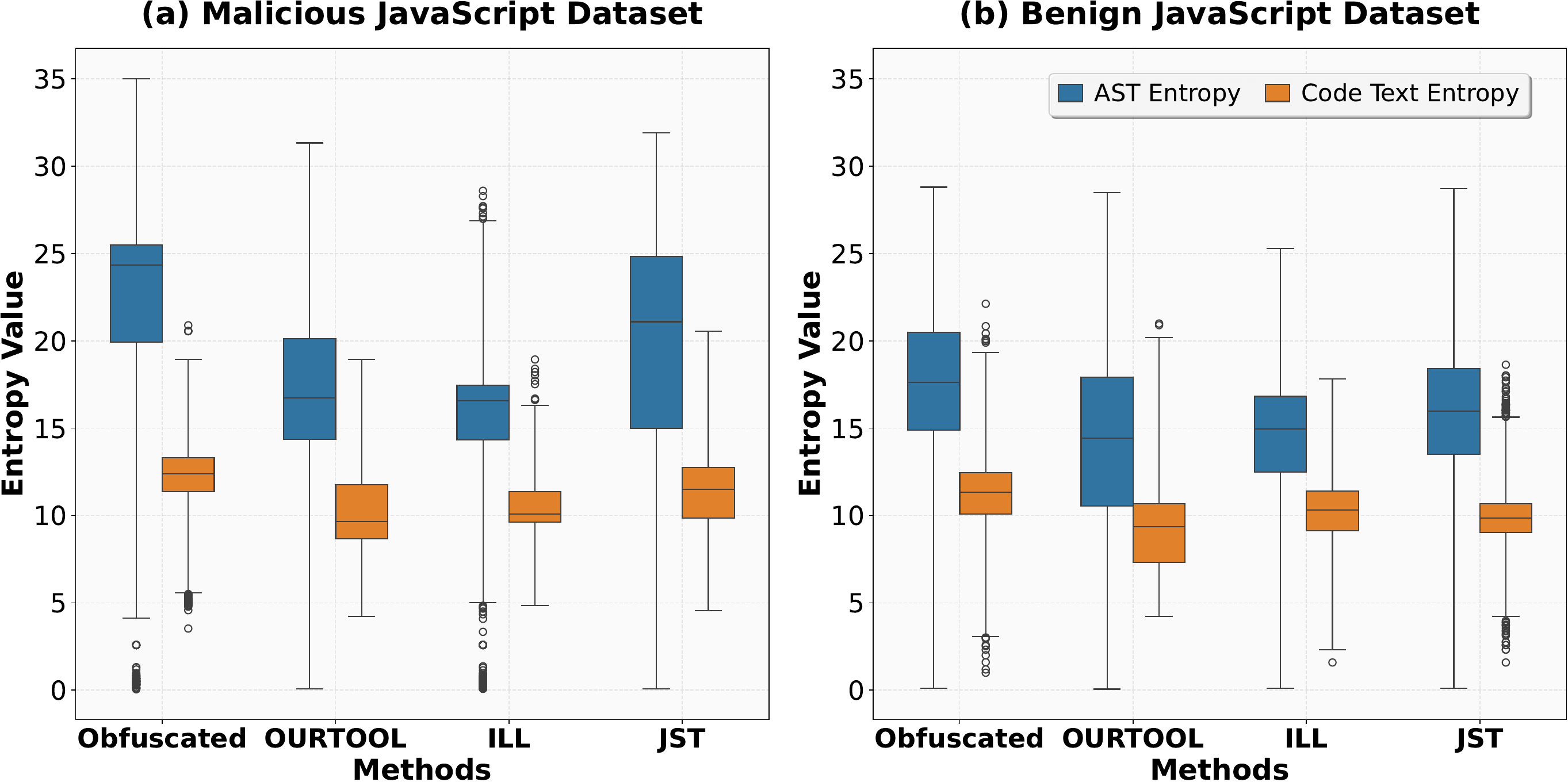}
\caption{Entropy reduction across (a) MalJS and (b) BenignJS.}
\label{fig:entropy_reduction_final}
\end{figure}

\noindent {\bf Effectiveness Results.} \OurTool successfully processed all 44,421 samples (100\%), while JST processed 30,188 samples (67.9\%) and ILL handled 14,140 samples (31.8\%). This comprehensive coverage is essential for practical malware analysis.
Figure~\ref{fig:entropy_reduction_final} shows \OurTool achieved the lowest median entropy values for both AST structure and code text across both datasets. The consistent reduction patterns between malicious and benign samples confirm our approach generalizes well beyond malware-specific obfuscation. The dual entropy reduction aligns with our HLR scores, confirming comprehensive deobfuscation capability. 

\noindent {\bf Performance Results.} Table~\ref{tab:efficiency} presents performance metrics showing \OurTool's 100\% success rate versus JST's 45.67\% and ILL's 19.06\%, with processing times of 83.17 seconds that are acceptable for malware analysis scenarios. \OurTool maintains consistent performance across all file sizes including very large files ($\textgreater10MB$) with moderate memory consumption (0.1034 MB average increment). In contrast, JST exhibits low memory usage (0.0056 MB) but poor success rate, while ILL's fixed allocation (0.3130 MB) limits capability to 7.58KB maximum file size, causing baseline tools to frequently timeout on complex samples.

\begin{table}[htbp]
\centering
\caption{Performance comparison across MalJS dataset.}
\label{tab:efficiency}
\begin{tabular}{lcccc}
\toprule
\textbf{Tool} & \textbf{Success} & \textbf{Time (s)} & \textbf{Memory (MB)} & \textbf{Max File} \\
\midrule
JST & 45.67\% & 22.69 & \textbf{+0.0056} & 17.51MB \\
ILL & 19.06\% & \textbf{6.26} & +0.3130 & 7.58KB \\
\OurTool & \textbf{100\%} & 83.17 & +0.1034 & \textbf{29.58MB} \\
\bottomrule
\end{tabular}
\end{table}

\subsection{LLM-based Realiability and Readability}
\label{subsec:llm_readability}
To evaluate practical effectiveness, we assess code readability improvement and processing reliability through LLM-based evaluation, avoiding human assessment subjectivity.
For readability assessment, we selected 60 sample pairs from MalJS for cross-model validation using Claude 3.7 Sonnet, Gemini 2.5 Pro, DeepSeek-R1, and GPT-o3. Each LLM scored obfuscated and deobfuscated code pairs on a 0-10 readability scale using customized prompts (Appendix~\ref{appendix:llm_prompt_score}). For reliability analysis, we selected 100 samples from Section~\ref{sec:ablation} for identifier renaming using GPT-4o-mini.

\noindent {\bf Readability Results.}
Table~\ref{tab:Avg_scores} shows substantial readability improvement across all evaluators. \OurTool achieves 466.94\% average improvement, transforming unreadable obfuscated code (scores 1.02-1.81) into comprehensible versions (scores 6.21-7.83). Cross-model consistency validates improvement objectivity. Deobfuscated scores around 6-8 points indicate practical readability for security analysis, corroborating HLR score 0.8820. Consistent improvements across samples support entropy reduction findings (detailed visualization in Appendix~\ref{appendix:readability_analysis}).

\begin{table}[htbp]
  \centering
  \caption{LLM-based readability assessment validating complexity reduction.}
  \label{tab:Avg_scores}
  \begin{tabular}{lp{1.7cm}p{1.7cm}p{1.7cm}}
  \toprule
    \textbf{Model} & \textbf{Original} & \textbf{Deobfuscated} & \textbf{Improvement} \\
    \midrule
    \rowcolor[HTML]{EFEFEF} Claude 3.7 Sonnet & 1.46 & 7.12 & +387.67\% \\
    Gemini 2.5 Pro & 1.02 & 7.33 & +618.63\% \\
    \rowcolor[HTML]{EFEFEF} DeepSeek-R1 & 1.09 & 7.83 & +618.35\% \\
    GPT-o3 & 1.81 & 6.21 & +243.09\% \\
    \bottomrule
  \end{tabular}
\end{table}



\noindent {\bf Realiability Results.}
The 100-sample GPT-4o-mini evaluation generated 20,897 API calls consuming 6,446,654 tokens, costing \$1.04 total with 10.96 hours processing time.  Average per-sample cost was \$0.0104 with 6.6 minutes processing time.  Long-tail distribution showed median cost \$0.0003 and median time 10.7 seconds.  Samples ranged from 250 bytes to 2.44 MB (average 80 KB, median 8 KB). Manual inspection revealed a 12\% hallucination rate, defined as identifier renaming deviating from ground truth expectations. Cases involved overly verbose naming (e.g., \texttt{i} to \texttt{iterationController}) reflecting over-interpretation. The remaining 88\% demonstrated reliable renaming.


\subsection{User Study}
To complement objective LLM-based assessment with human validation of real-world usability, we conducted a user study with 9 participants across three expertise levels: novice (3 participants: undergraduates and developers), intermediate (3 participants: graduate students and CTF players), and expert (3 participants: security researchers and analysts).
Each participant analyzed 6 obfuscated JavaScript samples under three conditions: Original (raw obfuscated), Traditional (processed by JST), and \OurTool. The samples were balanced across complexity levels and malicious behavior types using a Latin square design~\cite{kirk2012experimental} to control for learning effects and individual differences. Participants completed standardized analysis tasks (functionality identification, variable recognition, execution flow ordering, threat classification, IOC extraction) and provided subjective evaluations on 5-point Likert scales~\cite{likert1932technique} for readability, clarity, logicality, difficulty, and confidence. The complete questionnaire is provided in Appendix~\ref{appendix:questionnaire}.

\noindent \textbf{Results.} Table~\ref{tab:user_study_results} (Appendix~\ref{appendix:user_study_table}) shows \OurTool consistently improved analysis accuracy across all groups. Novices gained the most (+12.7\%). Analysis time decreased substantially, especially for intermediates (-47.7\%). Subjective ratings for readability, clarity, and logicality showed marked improvements, often doubling compared to Original. Higher difficulty and confidence scores indicate \OurTool improved code analyzability and boosted analysts' confidence.
Participant feedback on a heavily obfuscated jQuery UI sample\footnote{Sample MD5: 365df2394d7f7cf6ebdc7dfac6d2b3e6} illustrates practical impact. One intermediate participant noted: \textit{The variable renaming made it possible to follow the logic flow, and I could identify the suspicious network calls within minutes instead of struggling for the entire session.} This aligns with quantitative results showing 50\% analysis time reduction and substantial readability improvements.

\label{evaluation}

\section{Discussion}
\subsection{\OurTool in Security Analysis Workflows}

To further demonstrate \OurTool's practical value, we evaluated its performance on the JSFireTruck campaign sample that motivated this research (Listing~\ref{lst:jsfiretruck_multilayer} in Section~\ref{subsec:motivation}). This case study reveals how \OurTool addresses real-world challenges that existing tools cannot handle.

As shown in Listing~\ref{lst:complete_deobfuscation}, \OurTool successfully deobfuscates the complete multi-layer obfuscation chain that defeats existing tools. While specialized tools like unjsfuck only handle the basic layer, leaving intermediate output such as \texttt{\$[0] + \$[1].match(\$[2]|\$[3])} requiring manual analysis, \OurTool processes all obfuscation layers automatically. The tool transforms thousands of characters of JSFireTruck-encoded expressions into clear, readable code that directly exposes malicious behavior,
such as referrer-based victim filtering, iframe injection with CSS manipulation, and evidence removal. Security analysts can quickly understand attack intent without manual deobfuscation, demonstrating \OurTool's practical value for real-world security analysis.
\begin{lstlisting}[style=JavaScript, caption={\OurTool's deobfuscated output revealing malicious intent.}, label={lst:complete_deobfuscation}]
if (document.referrer.match(/google|bing|yahoo|duckduckgo/)) {
var maliciousIframe = '<iframe src="' + maliciousUrl + 
'" style="z-index:30000;width:100%;height:100%;' +
'position:absolute;top:0;left:0;"></iframe>';
document.getElementById(randomId).innerHTML = maliciousIframe;
}
document.currentScript.remove();
\end{lstlisting}

\subsection{Anti-Analysis Technique Handling}
\label{subsec:anti_analysis}

Beyond traditional obfuscation, malicious JavaScript employs anti-analysis techniques to evade detection. \OurTool handles these through static analysis and dynamic execution.

\noindent {\bf Time Bombs.} \OurTool combines static analysis for time-based logic simplification with controlled dynamic execution. Listing~\ref{lst:time} (Appendix~\ref{appendix:anti_analysis_examples}) shows time bomb handling where time-based conditions are exposed through static simplification, allowing analysts to identify activation timestamps (January 1, 2022 in this example) and assess threat timelines without triggering malicious payloads.

\noindent {\bf Environmental Probes.} \OurTool handles environment detection through browser API simulation and controlled sandbox execution. Our approach provides DOM-like stubs (\texttt{document}, \texttt{window}, \texttt{navigator}) and whitelists essential browser APIs (\texttt{btoa}, \texttt{atob}, \texttt{setTimeout}) as safe globals, preventing \texttt{ReferenceError} exceptions while ensuring environment checks receive plausible responses without exposing the host system. This allows deobfuscation to complete and expose evasion logic for analyst review.



\subsection{Implications for LLM-based Deobfuscation}
Our results have important implications for LLM-based approaches in deobfuscation workflows. While JsDeObsBench demonstrated potential of pure LLM approaches for code simplification and readability improvement, it revealed significant reliability concerns, context window constraints, unpredictable failure modes, and high computational costs that limit practical deployment.  Our hybrid approach validates the complementary nature of traditional and LLM-based methods by utilizing static AST analysis and dynamic execution for core deobfuscation tasks while leveraging LLM capabilities for readable output generation.  This targeted integration achieves both traditional method reliability and LLM readability improvements (466.94\% average improvement with 88\% processing reliability at \$0.0104 per sample).

\subsection{Limitations}
While \OurTool shows a significant advancement in JavaScript deobfuscation capabilities, some limitations remain. \looseness=-1

\noindent {\bf Compatibility Constraints.} 
To support older syntax, ES6+ features may be downgraded during preprocessing, potentially limiting modern JavaScript capabilities in deobfuscated output. We mitigate this through selective normalization disabling and multi-parser fallback mechanisms.

\noindent {\bf LLM Integration Challenges.} 
Our Humanizer module faces inherent challenges in understanding obfuscated code context and domain-specific terminology. LLMs may generate semantically plausible but contextually inappropriate names, particularly in malware analysis where technical precision is crucial. Context window limitations may prevent full understanding of complex variable relationships across large codebases. We address this through multi-provider LLM architecture for suggestion comparison, validation mechanisms checking generated names against existing scope bindings, and fallback strategies preserving original identifiers when validation fails. A detailed failure case is provided in Appendix~\ref{appendix:failure_case}.

\noindent {\bf Security and Scope Limitations.} 
\OurTool faces several constraints, including adversarial resistance and scope limitations. Sophisticated adversaries can craft targeted obfuscation to bypass specific techniques, creating an inevitable arms race. As a deobfuscator, we cannot handle cryptographic packing or custom language interpreters that hide code. Our approach also cannot handle time-dependent obfuscation techniques that use runtime values for code decryption. Our dynamic execution using Node.js \texttt{vm.runInNewContext} carries theoretical VM escape risks that may require additional containerization.

\label{discuss}

\section{related work}
\subsection{JavaScript Obfuscation Detection}
Large-scale analyses demonstrate obfuscation's widespread adoption~\cite{dsn21, imc20, jack2023, hidenoseek, wobfuscator, ieeeaccess19, caai20}. Moog~\cite{dsn21} analyzes Alexa Top 10K scripts using control-flow-enhanced AST classifiers, while Sarker~\cite{imc20} discovers most domains host obfuscated scripts through runtime-static analysis. Jack-in-the-box~\cite{jack2023} examines vulnerable npm bundles in top sites. Advanced evasion techniques continue emerging: HideNoSeek~\cite{hidenoseek} camouflages malware in benign ASTs, Wobfuscator~\cite{wobfuscator} embeds logic in WebAssembly, and JSPro~\cite{ieeeaccess19} reduces runtime overhead via WASM translation. Detection methods include runtime-static discrepancy analysis~\cite{imc20} and FastText with virtual machine-based deobfuscation~\cite{caai20}. 
While these studies focus on obfuscation detection, our work addresses the complementary challenge of comprehensive deobfuscation. 

\subsection{JavaScript Deobfuscation Techniques}
Researchers have developed various deobfuscation approaches to counter script obfuscation~\cite{jsimpo, transast, esecfse17, deobf2023, acmtops21, jsdes, scam20, smartgencon21}. Most techniques focus on specific obfuscation aspects. JSimpo~\cite{jsimpo} provides structural deobfuscation, TransAST~\cite{transast} leverages AST template mappings for identifier recovery, while JSNaughty~\cite{esecfse17} uses statistical machine translation and DEOBFUSCATING~\cite{deobf2023} employs character-based tokenization with Seq2Seq models. Eval obfuscation receives targeted solutions from Arceri~\cite{acmtops21} through abstract interpretation and JSDES~\cite{jsdes} via dynamic stubbing. Multi-technique approaches include SAFE-DEOBS~\cite{scam20} with AST-based optimization passes and Hajarnis~\cite{smartgencon21} integrating taint analysis and symbolic execution. Recent LLM evaluations reveal significant limitations: GPT-4 shows poor functional deobfuscation for advanced techniques~\cite{usenix24}, while JsDeObsBench~\cite{jsdeobsbench2025} demonstrates critical reliability issues with average failure rates of 2.76\% for syntax correctness and 37.40\% for execution correctness.
Unlike these fragmented approaches that excel only in narrow domains or suffer from reliability issues, \OurTool provides comprehensive deobfuscation through systematic integration of static AST analysis, controlled dynamic execution, and targeted LLM post-processing.

\subsection{Other Language Deobfuscation Research} 
Deobfuscation research across languages provides valuable insights~\cite{dobf, usenix24, powerpeeler, oblivion, api-xray, promba, mba-blast, gamba, stringhound, deprodexcator, sok, php2021}. LLM approaches show mixed results: DOBF~\cite{dobf} outperforms CodeBERT through pre-training, while GPT-4 achieves limited functional deobfuscation~\cite{usenix24}. Dynamic methods demonstrate effectiveness through PowerPeeler~\cite{powerpeeler} for PowerShell reconstruction, Oblivion~\cite{oblivion} for VBA macro instrumentation, and API-Xray~\cite{api-xray} for import reconstruction. Expression simplification techniques include ProMBA~\cite{promba} for MBA synthesis, MBA-Blast~\cite{mba-blast} with transformations, and GAMBA~\cite{gamba} for nonlinear MBAs. Language-specific tools target particular environments: StringHound~\cite{stringhound} for Android string obfuscation, DeProDexcator~\cite{deprodexcator} for ProGuard reversal, while systematic knowledge surveys~\cite{sok} analyze virtualization deobfuscation approaches, and PHP research~\cite{php2021} explores layout obfuscation with AES encryption. 
\OurTool is purpose-built for JavaScript's unique characteristics
to provide native handling of JavaScript-specific obfuscation patterns while maintaining compatibility with the language's execution model and ecosystem requirements.
\label{related}

\section{conclusion}
This work proposes \OurTool, a comprehensive JavaScript deobfuscation tool. \OurTool addresses limitations in existing approaches through multi-stage processing and LLM integration. It handles 20 categories of obfuscation techniques using static AST analysis, dynamic execution, and semantic-aware code humanization. We maintains correctness through strict semantic validation and CFG/DDG consistency checking. Experimental results demonstrate \OurTool's effectiveness across multiple evaluation dimensions. \OurTool successfully processes all samples in evaluation datasets. It maintains 100\% correctness on processed samples and achieves 83\% semantic consistency, outperforming both traditional and LLM-based approaches. Code simplification evaluation demonstrates an HLR score of 0.8820. LLM-based readability assessment shows 466.94\% average improvement. Large-scale evaluation on 44,421 samples demonstrates significant entropy reduction, confirming \OurTool's effectiveness for real-world malware analysis.
\label{conclusion}

\section*{Ethics Considerations}
This research follows established cybersecurity ethics. Benign samples come from public sources, and malicious samples from our corporate partner for defense only. Experiments were performed in isolated environments, and LLM evaluations used authorized APIs. Data handling ensures samples stay secure without redistribution. When releasing datasets, we will require identity verification and intent declaration to prevent misuse, balancing open science with necessary security safeguards.


%
\bibliographystyle{IEEEtran}
\bibliography{ref}

\begin{thebibliography}{10}
\providecommand{\url}[1]{#1}
\csname url@samestyle\endcsname
\providecommand{\newblock}{\relax}
\providecommand{\bibinfo}[2]{#2}
\providecommand{\BIBentrySTDinterwordspacing}{\spaceskip=0pt\relax}
\providecommand{\BIBentryALTinterwordstretchfactor}{4}
\providecommand{\BIBentryALTinterwordspacing}{\spaceskip=\fontdimen2\font plus
\BIBentryALTinterwordstretchfactor\fontdimen3\font minus \fontdimen4\font\relax}
\providecommand{\BIBforeignlanguage}[2]{{%
\expandafter\ifx\csname l@#1\endcsname\relax
\typeout{** WARNING: IEEEtran.bst: No hyphenation pattern has been}%
\typeout{** loaded for the language `#1'. Using the pattern for}%
\typeout{** the default language instead.}%
\else
\language=\csname l@#1\endcsname
\fi
#2}}
\providecommand{\BIBdecl}{\relax}
\BIBdecl

\bibitem{ren2023empirical}
K.~Ren, W.~Qiang, Y.~Wu, Y.~Zhou, D.~Zou, and H.~Jin, ``An empirical study on the effects of obfuscation on static machine learning-based malicious javascript detectors,'' in \emph{Proceedings of the 32nd ACM SIGSOFT International Symposium on Software Testing and Analysis (ISSTA)}, 2023.

\bibitem{ndichu2020deobfuscation}
S.~Ndichu, S.~Kim, and S.~Ozawa, ``Deobfuscation, unpacking, and decoding of obfuscated malicious javascript for machine learning models detection performance improvement,'' \emph{CAAI Transactions on Intelligence Technology}, 2020.

\bibitem{gong2018dynamic}
L.~Gong, ``Dynamic analysis for javascript code,'' Ph.D. dissertation, University of California, Berkeley, 2018.

\bibitem{vilchik2024static}
\BIBentryALTinterwordspacing
E.~Vilchik, ``Static analysis in javascript: What's easy and what's hard,'' GitNation Conference Talk, 2024, conference presentation, Accessed: Jan. 15, 2025. [Online]. Available: \url{https://gitnation.com/contents/static-analysis-in-javascript-whats-easy-and-whats-hard}
\BIBentrySTDinterwordspacing

\bibitem{trendmicro:socgholish2025}
\BIBentryALTinterwordspacing
{Trend Micro Research}, ``{SocGholish}'s intrusion techniques facilitate distribution of ransomware,'' Trend Micro Research, 2025, accessed: Mar. 27, 2025. [Online]. Available: \url{https://www.trendmicro.com/en_us/research/25/c/socgholishs-intrusion-techniques-facilitate-distribution-of-rans.html}
\BIBentrySTDinterwordspacing

\bibitem{proofpoint2022socgholish}
\BIBentryALTinterwordspacing
P.~T.~R. Team, ``Part 1: Socgholish: A very real threat from a very fake update,'' December 2022, accessed: Jan. 15, 2025. [Online]. Available: \url{https://www.proofpoint.com/us/blog/threat-insight/part-1-socgholish-very-real-threat-very-fake-update}
\BIBentrySTDinterwordspacing

\bibitem{brezinski2023metamorphic}
\BIBentryALTinterwordspacing
K.~Brezinski and K.~Ferens, ``Metamorphic malware and obfuscation: A survey of techniques, variants, and generation kits,'' Security and Communication Networks, Wiley, 2023, article ID: 8227751, Accessed: Jan. 15, 2025. [Online]. Available: \url{https://onlinelibrary.wiley.com/doi/10.1155/2023/8227751}
\BIBentrySTDinterwordspacing

\bibitem{jstillery}
\BIBentryALTinterwordspacing
{Minded Security}, ``{JStillery} - advanced {JavaScript} deobfuscation via partial evaluation,'' GitHub Repository, 2024, accessed: Jul. 3, 2025. [Online]. Available: \url{https://github.com/mindedsecurity/jstillery}
\BIBentrySTDinterwordspacing

\bibitem{illuminatejs}
\BIBentryALTinterwordspacing
{Geeks on Security}, ``{IlluminateJS} - a static {JavaScript} deobfuscator,'' GitHub Repository, 2024, accessed: Jul. 3, 2025. [Online]. Available: \url{https://github.com/geeksonsecurity/illuminatejs}
\BIBentrySTDinterwordspacing

\bibitem{raychev2015predicting}
V.~Raychev, M.~Vechev, and A.~Krause, ``Predicting program properties from big code,'' in \emph{Proceedings of the 42nd ACM SIGPLAN-SIGACT Symposium on Principles of Programming Languages (POPL)}, 2015.

\bibitem{kasada2023javascript}
\BIBentryALTinterwordspacing
Kasada, ``Javascript deobfuscation: Hiding intent to fortify bot defenses,'' Kasada Security Blog, 2023, accessed: Jan. 15, 2025. [Online]. Available: \url{https://www.kasada.io/javascript-deobsfusction-bot-defenses/}
\BIBentrySTDinterwordspacing

\bibitem{lielmanis2024jsbeautify}
\BIBentryALTinterwordspacing
E.~Lielmanis and L.~Newman, ``js-beautify: Beautifier for javascript, html, css,'' GitHub Repository, 2024, accessed: Jan. 15, 2025. [Online]. Available: \url{https://github.com/beautifier/js-beautify}
\BIBentrySTDinterwordspacing

\bibitem{synchrony2024}
\BIBentryALTinterwordspacing
Relative, ``Synchrony: Javascript cleaner \& deobfuscator,'' GitHub Repository, 2024, accessed: Jan. 15, 2025. [Online]. Available: \url{https://github.com/relative/synchrony}
\BIBentrySTDinterwordspacing

\bibitem{bensb_deobfuscator_io}
\BIBentryALTinterwordspacing
{ben-sb}, ``deobfuscator-io-archived: Deobfuscator for obfuscator.io,'' GitHub Repository, 2024, accessed: Jan. 15, 2025. [Online]. Available: \url{https://github.com/ben-sb/javascript-deobfuscator}
\BIBentrySTDinterwordspacing

\bibitem{jsdeobsbench2025}
G.~Chen, X.~Jin, and Z.~Lin, ``Jsdeobsbench: Measuring and benchmarking llms for javascript deobfuscation,'' in \emph{Proceedings of the 2025 ACM SIGSAC Conference on Computer and Communications Security (CCS)}, 2025.

\bibitem{unit42_jsfiretruck}
\BIBentryALTinterwordspacing
{Unit 42}, ``{JSFireTruck}: Exploring malicious {JavaScript} using {JSF*ck} as an obfuscation technique,'' Palo Alto Networks Unit 42, 2025, accessed: Jul. 3, 2025. [Online]. Available: \url{https://unit42.paloaltonetworks.com/malicious-javascript-using-jsfiretruck-as-obfuscation/}
\BIBentrySTDinterwordspacing

\bibitem{jsimplifier}
\BIBentryALTinterwordspacing
{XingTuLab}, ``{JSIMPLIFIER} – {JavaScript} code simplification and deobfuscation tool,'' GitHub Repository, 2025, accessed: Dec, 2025. [Online]. Available: \url{https://github.com/XingTuLab/JSIMPLIFIER}
\BIBentrySTDinterwordspacing

\bibitem{ecmascript2023}
\BIBentryALTinterwordspacing
{Ecma International}, ``Ecmascript 2023 language specification,'' Ecma International, Standard ECMA-262, 2023, accessed: Jan. 15, 2025. [Online]. Available: \url{https://www.ecma-international.org/publications-and-standards/standards/ecma-262/}
\BIBentrySTDinterwordspacing

\bibitem{jsf*ck:handwiki}
\BIBentryALTinterwordspacing
{HandWiki}, ``{JSF*ck},'' HandWiki Encyclopedia, 2024, accessed: Dec. 19, 2024. [Online]. Available: \url{https://handwiki.org/wiki/JSFuck}
\BIBentrySTDinterwordspacing

\bibitem{skolka2019anything}
P.~Skolka, C.-A. Staicu, and M.~Pradel, ``Anything to hide? studying minified and obfuscated code in the web,'' in \emph{The World Wide Web Conference (WWW)}, 2019.

\bibitem{xu2013power}
W.~Xu, F.~Zhang, and S.~Zhu, ``The power of obfuscation techniques in malicious javascript code: A measurement study,'' in \emph{International Conference on Malicious and Unwanted Software: "The Americas" (MALWARE)}, 2013.

\bibitem{dsn21}
M.~Moog, M.~Demmel, M.~Backes, and A.~Fass, ``Statically detecting {JavaScript} obfuscation and minification techniques in the wild,'' in \emph{Proceedings of the IEEE/IFIP International Conference on Dependable Systems and Networks (DSN)}, 2021.

\bibitem{imc20}
S.~Sarker, J.~Jueckstock, and A.~Kapravelos, ``Hiding in plain site: Detecting {JavaScript} obfuscation through concealed browser {API} usage,'' in \emph{Proceedings of the ACM Internet Measurement Conference (IMC)}, 2020.

\bibitem{jack2023}
J.~Rack and C.-A. Staicu, ``Jack-in-the-box: An empirical study of {JavaScript} bundling on the web and its security implications,'' in \emph{Proceedings of the ACM SIGSAC Conference on Computer and Communications Security (CCS)}, 2023.

\bibitem{hidenoseek}
A.~Fass, M.~Backes, and B.~Stock, ``{HideNoSeek}: Camouflaging malicious {JavaScript} in benign {ASTs},'' in \emph{Proceedings of the ACM SIGSAC Conference on Computer and Communications Security (CCS)}, 2019.

\bibitem{wobfuscator}
A.~Romano, Y.~Zheng, W.~Wang, L.~Ying, and G.~Vigna, ``{Wobfuscator}: Obfuscating {JavaScript} malware via opportunistic translation to {WebAssembly},'' in \emph{Proceedings of the IEEE Symposium on Security and Privacy (S\&P)}, 2022.

\bibitem{ieeeaccess19}
S.~Wang, P.~Wang, and D.~Wu, ``Leveraging {WebAssembly} for numerical {JavaScript} code virtualization,'' \emph{IEEE Access}, 2019.

\bibitem{caai20}
S.~Ndichu, S.~Kim, and S.~Ozawa, ``Deobfuscation, unpacking, and decoding of obfuscated malicious {JavaScript} for machine learning models detection performance improvement,'' \emph{CAAI Transactions on Intelligence Technology}, 2020.

\bibitem{jscrambler}
\BIBentryALTinterwordspacing
{Jscrambler}, ``Pioneering client-side protection platform,'' Jscrambler, 2024, accessed: Jul. 3, 2025. [Online]. Available: \url{https://jscrambler.com/}
\BIBentrySTDinterwordspacing

\bibitem{jshaman}
\BIBentryALTinterwordspacing
{JShaman}, ``{JavaScript} obfuscation encryption,'' Shaman Technology, 2024, accessed: Jul. 3, 2025. [Online]. Available: \url{https://www.jshaman.com/}
\BIBentrySTDinterwordspacing

\bibitem{metasploit_js_obfuscation}
\BIBentryALTinterwordspacing
{Metasploit Documentation}, ``{JavaScript} obfuscation,'' Rapid7 Metasploit Documentation, 2024, accessed: Jul. 3, 2025. [Online]. Available: \url{https://docs.metasploit.com/docs/development/developing-modules/libraries/obfuscation/how-to-obfuscate-javascript-in-metasploit.html}
\BIBentrySTDinterwordspacing

\bibitem{uglifyjs}
\BIBentryALTinterwordspacing
{Mishoo}, ``{UglifyJS} - {JavaScript} parser / mangler / compressor / beautifier toolkit,'' GitHub Repository, 2024, accessed: Jul. 3, 2025. [Online]. Available: \url{https://github.com/mishoo/UglifyJS}
\BIBentrySTDinterwordspacing

\bibitem{ob}
\BIBentryALTinterwordspacing
{JavaScript Obfuscator Tool}, ``A free and efficient obfuscator for {JavaScript},'' obfuscator.io, 2024, accessed: Jul. 3, 2025. [Online]. Available: \url{https://obfuscator.io/}
\BIBentrySTDinterwordspacing

\bibitem{jsjiami}
\BIBentryALTinterwordspacing
{JSJiami}, ``Online {JavaScript} code obfuscation and encryption tool,'' jsjiami.com, 2024, accessed: Jul. 3, 2025. [Online]. Available: \url{https://www.jsjiami.com/}
\BIBentrySTDinterwordspacing

\bibitem{daft}
\BIBentryALTinterwordspacing
{Daft Logic}, ``Online {Javascript} obfuscator,'' Daft Logic, 2024, accessed: Jul. 3, 2025. [Online]. Available: \url{https://www.daftlogic.com/projects-online-javascript-obfuscator.htm}
\BIBentrySTDinterwordspacing

\bibitem{terser}
\BIBentryALTinterwordspacing
{Nodejs.cn}, ``{Terser} - {JavaScript} obfuscator and compressor toolkit,'' Terser Official Documentation, 2024, accessed: Jul. 3, 2025. [Online]. Available: \url{https://terser.nodejs.cn/}
\BIBentrySTDinterwordspacing

\bibitem{free}
\BIBentryALTinterwordspacing
{FreeJSObfuscator Team}, ``Free {JavaScript} obfuscator - 100\% free {JS} code protection tool,'' FreeJSObfuscator.com, 2024, accessed: Jul. 3, 2025. [Online]. Available: \url{https://www.freejsobfuscator.com/}
\BIBentrySTDinterwordspacing

\bibitem{stunnix}
\BIBentryALTinterwordspacing
{Stunnix}, ``{JavaScript} obfuscator - protect {JavaScript} code from illegal reuse and ip theft,'' Stunnix, 2024, accessed: Jul. 3, 2025. [Online]. Available: \url{http://stunnix.com/prod/jo/}
\BIBentrySTDinterwordspacing

\bibitem{gnirts}
\BIBentryALTinterwordspacing
{Anseki}, ``{gnirts} - obfuscate string literals in {JavaScript} code,'' GitHub Pages, 2024, accessed: Jul. 3, 2025. [Online]. Available: \url{https://anseki.github.io/gnirts/}
\BIBentrySTDinterwordspacing

\bibitem{beautify}
\BIBentryALTinterwordspacing
{BeautifyTools}, ``Online {Javascript} obfuscator - protect {JavaScript} code from theft,'' BeautifyTools, 2024, accessed: Jul. 3, 2025. [Online]. Available: \url{https://beautifytools.com/javascript-obfuscator.php}
\BIBentrySTDinterwordspacing

\bibitem{unjsfuck}
\BIBentryALTinterwordspacing
{karust}, ``{unjsfuck} - {JSFuck} deobfuscator,'' GitHub, 2024, accessed: Jul. 4, 2025. [Online]. Available: \url{https://github.com/karust/unjsfuck}
\BIBentrySTDinterwordspacing

\bibitem{meriyah}
\BIBentryALTinterwordspacing
{Meriyah Team}, ``{Meriyah} - a 100\% compliant, self-hosted {javascript} parser,'' GitHub Repository, 2024, accessed: Jul. 3, 2025. [Online]. Available: \url{https://github.com/meriyah/meriyah}
\BIBentrySTDinterwordspacing

\bibitem{webcrack}
\BIBentryALTinterwordspacing
{j4k0xb}, ``{webcrack} - deobfuscate obfuscator.io, unminify and unpack bundled {javascript},'' GitHub Repository, 2024, accessed: Jul. 3, 2025. [Online]. Available: \url{https://github.com/j4k0xb/webcrack}
\BIBentrySTDinterwordspacing

\bibitem{webpack}
\BIBentryALTinterwordspacing
{Webpack Team}, ``{webpack} - a static module bundler for modern {JavaScript} applications,'' webpack.js.org, 2024, accessed: Jul. 3, 2025. [Online]. Available: \url{https://webpack.js.org/}
\BIBentrySTDinterwordspacing

\bibitem{browserify}
\BIBentryALTinterwordspacing
{Browserify Team}, ``{Browserify} - browser-side require() the node.js way,'' browserify.org, 2024, accessed: Jul. 3, 2025. [Online]. Available: \url{https://browserify.org/}
\BIBentrySTDinterwordspacing

\bibitem{esprima}
\BIBentryALTinterwordspacing
A.~Hidayat, ``{Esprima} - {ECMAScript} parsing infrastructure for multipurpose analysis,'' Esprima.org, 2024, accessed: Jul. 3, 2025. [Online]. Available: \url{https://esprima.org/}
\BIBentrySTDinterwordspacing

\bibitem{acorn}
\BIBentryALTinterwordspacing
{Acorn Team}, ``{Acorn} - a tiny, fast {JavaScript} parser, written completely in {JavaScript},'' GitHub Repository, 2024, accessed: Jul. 3, 2025. [Online]. Available: \url{https://github.com/acornjs/acorn}
\BIBentrySTDinterwordspacing

\bibitem{babel}
\BIBentryALTinterwordspacing
{Babel Team}, ``{Babel} - {JavaScript} compiler,'' Babel Official Website, 2024, accessed: Jul. 4, 2025. [Online]. Available: \url{https://babeljs.io/}
\BIBentrySTDinterwordspacing

\bibitem{kolbitsch2012rozzle}
C.~Kolbitsch, B.~Livshits, B.~Zorn, and C.~Seifert, ``Rozzle: De-cloaking internet malware,'' ser. SP '12.\hskip 1em plus 0.5em minus 0.4em\relax USA: IEEE Computer Society, 2012.

\bibitem{humanify}
\BIBentryALTinterwordspacing
{Jehna}, ``{Humanify} - deobfuscate {Javascript} code using {ChatGPT},'' GitHub Repository, 2024, accessed: Jul. 3, 2025. [Online]. Available: \url{https://github.com/jehna/humanify}
\BIBentrySTDinterwordspacing

\bibitem{prettier}
\BIBentryALTinterwordspacing
{Prettier Team}, ``{Prettier} - opinionated code formatter,'' prettier.io, 2024, accessed: Jul. 3, 2025. [Online]. Available: \url{https://prettier.io/docs/}
\BIBentrySTDinterwordspacing

\bibitem{chai2022invoke}
H.~Chai, L.~Ying, H.~Duan, and D.~Zha, ``Invoke-deobfuscation: Ast-based and semantics-preserving deobfuscation for powershell scripts,'' in \emph{Proceedings of the 52nd Annual IEEE/IFIP International Conference on Dependable Systems and Networks (DSN)}, 2022.

\bibitem{tranco2024}
\BIBentryALTinterwordspacing
{Tranco Team}, ``Tranco: A research-oriented top sites ranking hardened against manipulation,'' 2024, accessed: 2024-12-19. [Online]. Available: \url{https://tranco-list.eu/}
\BIBentrySTDinterwordspacing

\bibitem{publicwww}
\BIBentryALTinterwordspacing
{PublicWWW}, ``{PublicWWW} - source code search engine,'' publicwww.com, 2024, accessed: Jul. 3, 2025. [Online]. Available: \url{https://publicwww.com/}
\BIBentrySTDinterwordspacing

\bibitem{devcoco2024jsdeobfuscator}
\BIBentryALTinterwordspacing
{dev-coco}, ``Javascript deobfuscator - online tool,'' Online Tool, 2024, accessed: Jan. 15, 2025. [Online]. Available: \url{https://dev-coco.github.io/Online-Tools/JavaScript-Deobfuscator.html}
\BIBentrySTDinterwordspacing

\bibitem{kuizuo2024jsdeobfuscator}
\BIBentryALTinterwordspacing
kuizuo, ``js-deobfuscator: Javascript code deobfuscation tool,'' GitHub Repository, 2024, accessed: Jan. 15, 2025. [Online]. Available: \url{https://github.com/kuizuo/js-deobfuscator}
\BIBentrySTDinterwordspacing

\bibitem{hax0r31337_jsdec}
\BIBentryALTinterwordspacing
{hax0r31337}, ``Jsdec: Online javascript decoder,'' GitHub Repository, 2021, supports sojson v4/Premium/v5. Project archived October 10, 2021, Accessed: Jan. 15, 2025. [Online]. Available: \url{https://github.com/hax0r31337/JSDec}
\BIBentrySTDinterwordspacing

\bibitem{jsimpo}
T.~Chen, D.~Li, Y.~Zhang, and T.~Xie, ``{JSimpo}: Structural deobfuscation of {JavaScript} programs,'' \emph{ACM Transactions on Software Engineering and Methodology (TOSEM)}, 2025.

\bibitem{aiohttp2025}
\BIBentryALTinterwordspacing
{aio-libs team}, ``aiohttp: Asynchronous http client/server framework for asyncio and python,'' GitHub Repository, 2025, version 3.12, Accessed: Jan. 15, 2025. [Online]. Available: \url{https://github.com/aio-libs/aiohttp}
\BIBentrySTDinterwordspacing

\bibitem{halstead1977}
M.~H. Halstead, \emph{Elements of Software Science}, ser. Operating and Programming Systems Series.\hskip 1em plus 0.5em minus 0.4em\relax Elsevier Science Inc, 1977.

\bibitem{kirk2012experimental}
R.~Kirk, \emph{Kirk, Roger E. (2013). Experimental design: Procedures for the behavioral sciences (4th ed.). Thousand Oaks, CA: Sage.}, 06 2013.

\bibitem{likert1932technique}
R.~Likert, ``A technique for the measurement of attitudes,'' \emph{Archives of psychology}, vol.~22, no. 140, p.~55, 1932.

\bibitem{transast}
Y.~Qin, W.~Wang, Z.~Chen, H.~Song, and S.~Zhang, ``{TransAST}: A machine translation-based approach for obfuscated malicious {JavaScript} detection,'' in \emph{Proceedings of the IEEE/IFIP International Conference on Dependable Systems and Networks (DSN)}, 2023.

\bibitem{esecfse17}
B.~Vasilescu, C.~Casalnuovo, and P.~Devanbu, ``Recovering clear, natural identifiers from obfuscated {JS} names,'' in \emph{Proceedings of the 2017 11th Joint Meeting on Foundations of Software Engineering (ESEC/FSE)}, 2017.

\bibitem{deobf2023}
A.-G. Sîrbu, ``Deobfuscating {JavaScript} code using character-based tokenization,'' \emph{Studia Universitatis Babeș-Bolyai Informatica}, 2023.

\bibitem{acmtops21}
V.~Arceri and I.~Mastroeni, ``Analyzing dynamic code: A sound abstract interpreter for evil eval,'' \emph{ACM Transactions on Privacy and Security (ACMTOPS)}, 2021.

\bibitem{jsdes}
M.~AbdelKhalek and A.~Shosha, ``{JSDES}: An automated de-obfuscation system for malicious {JavaScript},'' in \emph{Proceedings of the 12th International Conference on Availability, Reliability and Security (ARES)}, 2017.

\bibitem{scam20}
A.~Herrera, ``Optimizing away {JavaScript} obfuscation,'' in \emph{Proceedings of the IEEE International Working Conference on Source Code Analysis and Manipulation (SCAM)}, 2020.

\bibitem{smartgencon21}
K.~Hajarnis, V.~Kshirsagar, S.~Doiphode, and A.~Joshi, ``A comprehensive solution for obfuscation detection and removal based on comparative analysis of deobfuscation tools,'' in \emph{Proceedings of the International Conference on Smart Generation Computing, Communication and Networking (SMART GENCON)}, 2021.

\bibitem{usenix24}
C.~Fang, Z.~Liu, Y.~Shi, J.~Huang, and Q.~Chen, ``Large language models for code analysis: Do {LLMs} really do their job?'' in \emph{Proceedings of the 33rd USENIX Security Symposium (USENIX Security)}, 2024.

\bibitem{dobf}
M.-A. Lachaux, B.~Roziere, M.~Szafraniec, and G.~Lample, ``{DOBF}: A deobfuscation pre-training objective for programming languages,'' in \emph{Proceedings of the 35th International Conference on Neural Information Processing Systems (NeurIPS)}, 2021.

\bibitem{powerpeeler}
R.~Li, C.~Zhang, H.~Chai, L.~Ying, H.~Duan, and J.~Tao, ``{PowerPeeler}: A precise and general dynamic deobfuscation method for {PowerShell} scripts,'' in \emph{Proceedings of the ACM SIGSAC Conference on Computer and Communications Security (CCS)}, 2024.

\bibitem{oblivion}
A.~Sanna, F.~Cara, D.~Maiorca, and G.~Giacinto, ``{Oblivion}: An open-source system for large-scale analysis of macro-based office malware,'' \emph{Journal of Computer Virology and Hacking Techniques}, 2024.

\bibitem{api-xray}
B.~Cheng, J.~Ming, J.~Burket, G.~Karame, and T.~Dumitraş, ``Obfuscation-resilient executable payload extraction from packed malware,'' in \emph{Proceedings of the 30th USENIX Security Symposium}, 2021.

\bibitem{promba}
J.~Lee and W.~Lee, ``Simplifying mixed boolean-arithmetic obfuscation by program synthesis and term rewriting,'' in \emph{Proceedings of the ACM SIGSAC Conference on Computer and Communications Security (CCS)}, 2023.

\bibitem{mba-blast}
B.~Liu, J.~Shen, J.~Ming, Q.~Zheng, J.~Li, and D.~Xu, ``{MBA-Blast}: Unveiling and simplifying mixed {Boolean-Arithmetic} obfuscation,'' in \emph{Proceedings of the 30th USENIX Security Symposium}, 2021.

\bibitem{gamba}
B.~Reichenwallner and P.~Meerwald-Stadler, ``Simplification of general mixed {Boolean-Arithmetic} expressions: {GAMBA},'' in \emph{Proceedings of the 2023 IEEE European Symposium on Security and Privacy Workshops (EuroS\&PW)}, 2023.

\bibitem{stringhound}
L.~Glanz, P.~Müller, L.~Baumgärtner, M.~Reif, S.~Amann, P.~Anthonysamy, and M.~Mezini, ``Hidden in plain sight: Obfuscated strings threatening your privacy,'' in \emph{Proceedings of the 15th ACM Asia Conference on Computer and Communications Security (ASIA CCS)}, 2020.

\bibitem{deprodexcator}
\BIBentryALTinterwordspacing
A.~Ashraaf, H.~Sarwar, G.~Murtaza, and M.~Ali, ``{DEPRODEXCATOR}: A tool for decompiling and deobfuscating {ProGuard} and {DexGuard} obfuscated android applications,'' engrXiv preprint, 2023, accessed: Jul. 3, 2025. [Online]. Available: \url{https://engrxiv.org/preprint/view/2910/5390}
\BIBentrySTDinterwordspacing

\bibitem{sok}
P.~Kochberger, S.~Schrittwieser, S.~Schweighofer, P.~Kieseberg, and E.~Weippl, ``{SoK}: Automatic deobfuscation of virtualization-protected applications,'' in \emph{Proceedings of the 16th International Conference on Availability, Reliability and Security (ARES)}, 2021.

\bibitem{php2021}
I.~Khairunisa and H.~Kabetta, ``{PHP} source code protection using layout obfuscation and {AES-256} encryption algorithm,'' in \emph{Proceedings of the International Workshop on Big Data and Information Security (IWBIS)}, 2021.

\bibitem{unit42:llm_obfuscation}
\BIBentryALTinterwordspacing
{Unit 42}, ``Now you see me, now you don't: Using {LLMs} to obfuscate malicious {JavaScript},'' Palo Alto Networks Unit 42, 2024, accessed: Dec. 27, 2024. [Online]. Available: \url{https://unit42.paloaltonetworks.com/using-llms-obfuscate-malicious-javascript/}
\BIBentrySTDinterwordspacing

\end{thebibliography}

\clearpage
\appendices

\section{Obfuscation Capability of Different Tools}
\label{appendix:ob_cab}
Table~\ref{tab:compare_obfuscation_tools} presents a comprehensive comparison of these tools' capabilities across all 20 obfuscation techniques.
\begin{table*}[htbp]
  \centering
  \begin{threeparttable}
  \caption{Comprehensive comparison of obfuscation capability of different tools.}
  \label{tab:compare_obfuscation_tools}
  \begin{tabular}{p{1.5cm}p{2.1cm}p{0.6cm}p{0.6cm}p{0.6cm}p{0.6cm}p{0.6cm}p{0.6cm}p{0.6cm}p{0.6cm}p{0.6cm}p{0.6cm}p{0.6cm}p{0.6cm}}
  \toprule
    \textbf{Type} & \textbf{Method} & \textbf{MeSpl}& \textbf{JSha}& \textbf{JSC}& \textbf{UGL}& \textbf{OBfu}& \textbf{JSV7}& \textbf{DAFT}& \textbf{TER}& \textbf{FRE}& \textbf{STU}& \textbf{GNI}& \textbf{BEA}\\
    \hline
    \multirow{5}{*}{\centering Lexical} & T0:Rename & $\checkmark$ & $\checkmark$ & $\checkmark$ & $\checkmark$ & $\checkmark$ & $\checkmark$ & $\checkmark$ & $\checkmark$ & $\checkmark$ & $\checkmark$ & $\checkmark$ & $\checkmark$\\
    \rowcolor[HTML]{EFEFEF}\cellcolor{white}                        & T1:Indirect &  & $\checkmark$ & $\checkmark$ &  & $\checkmark$ & $\checkmark$ & $\checkmark$ &  & $\checkmark$ & $\checkmark$ & $\checkmark$ & \\
                            & T2:Arithmetize & $\checkmark$ & $\checkmark$ &  &  & $\checkmark$ &  &  &  & $\checkmark$ & $\checkmark$ &  & \\
    \rowcolor[HTML]{EFEFEF}\cellcolor{white}                        & T3:StringEncode & $\checkmark$ & $\checkmark$ & $\checkmark$ &  & $\checkmark$ & $\checkmark$ &  & $\checkmark$ & $\checkmark$ & $\checkmark$ & $\checkmark$ & \\
                            & T4:BooleanEncode &  & $\checkmark$ & $\checkmark$ &  &  &  &  & $\checkmark$ &  &  &  & \\
    \hline
    \rowcolor[HTML]{EFEFEF}\cellcolor{white}\multirow{6}{*}[1.5em]{\centering Syntactic} & T5:Expr2Function &  & $\checkmark$ &  &  &  &  & $\checkmark$ &  &  &  &  & $\checkmark$\\
                            & T6:Assign2Function & $\checkmark$ & $\checkmark$ & $\checkmark$ &  &  & $\checkmark$ &  &  & $\checkmark$ & $\checkmark$ &  & \\
    \rowcolor[HTML]{EFEFEF}\cellcolor{white}                        & T7:Reverse &  & $\checkmark$ &  &  & $\checkmark$ &  &  &  &  &  &  & \\
                            & T8:AAEncode &  &  &  &  &  &  &  &  &  &  &  &  \\
    \rowcolor[HTML]{EFEFEF}\cellcolor{white}                        & T9:JJEncode &  &  &  &  &  &  &  &  &  &  &  &  \\
                          & T10:JSFUCK &  &  &  &  &  &  &  &  &  &  &  &  \\
    \hline
    \rowcolor[HTML]{EFEFEF}\cellcolor{white}\multirow{7}{*}{\centering Semantic} & T11:Arrayize & $\checkmark$ & $\checkmark$ & $\checkmark$ &  & $\checkmark$ &  &  & $\checkmark$ & $\checkmark$ & $\checkmark$ &  & \\
                            & T12:StrArrEncode &  & $\checkmark$ & $\checkmark$ &  & $\checkmark$ &  &  & $\checkmark$ & $\checkmark$ & $\checkmark$ &  & \\
    \rowcolor[HTML]{EFEFEF}\cellcolor{white}                        & T13:JSONEncode &  & $\checkmark$ &  &  &  &  &  &  &  &  &  &  \\
                            & T14:RegexpEncode &  & $\checkmark$ & $\checkmark$ &  &  & $\checkmark$ &  &  &  &  &  & $\checkmark$\\
    \rowcolor[HTML]{EFEFEF}\cellcolor{white}                        & T15:Eval &  & $\checkmark$ &  &  &  &  & $\checkmark$ &  &  & $\checkmark$ &  & $\checkmark$\\
                            & T16:Flattern & $\checkmark$ & $\checkmark$ & $\checkmark$ &  & $\checkmark$ &  &  &  &  &  &  & \\
    \rowcolor[HTML]{EFEFEF}\cellcolor{white}                        & T17:Insert &  & $\checkmark$ & $\checkmark$ &  & $\checkmark$ & $\checkmark$ &  &  &  &  &  & \\
    \hline
    \multirow{2}{*}[2pt]{\centering Multi-Layered} & T18:OB &  &  &  &  & $\checkmark$ &  &  &  &  &  &  & \\
    \rowcolor[HTML]{EFEFEF}\cellcolor{white}                        & T19:LLM & $\circ$ & $\checkmark$ & $\checkmark$ &  & $\checkmark$ & $\checkmark$ &  &  & $\checkmark$ & $\checkmark$ &  & \\
    \bottomrule
  \end{tabular}
  \begin{tablenotes}
    \footnotesize
    \item Note: $\checkmark$ = fully successful, $\circ$ = partially successful, blank = fully unsupported.
    \end{tablenotes}
  \end{threeparttable}
\end{table*}
\section{CFG/DDG Validation Example}
\label{appendix:cfg_ddg}
\begin{figure*}[htbp]
\centering
    \centering
    \includegraphics[width=\textwidth, height=0.4\textheight, keepaspectratio]{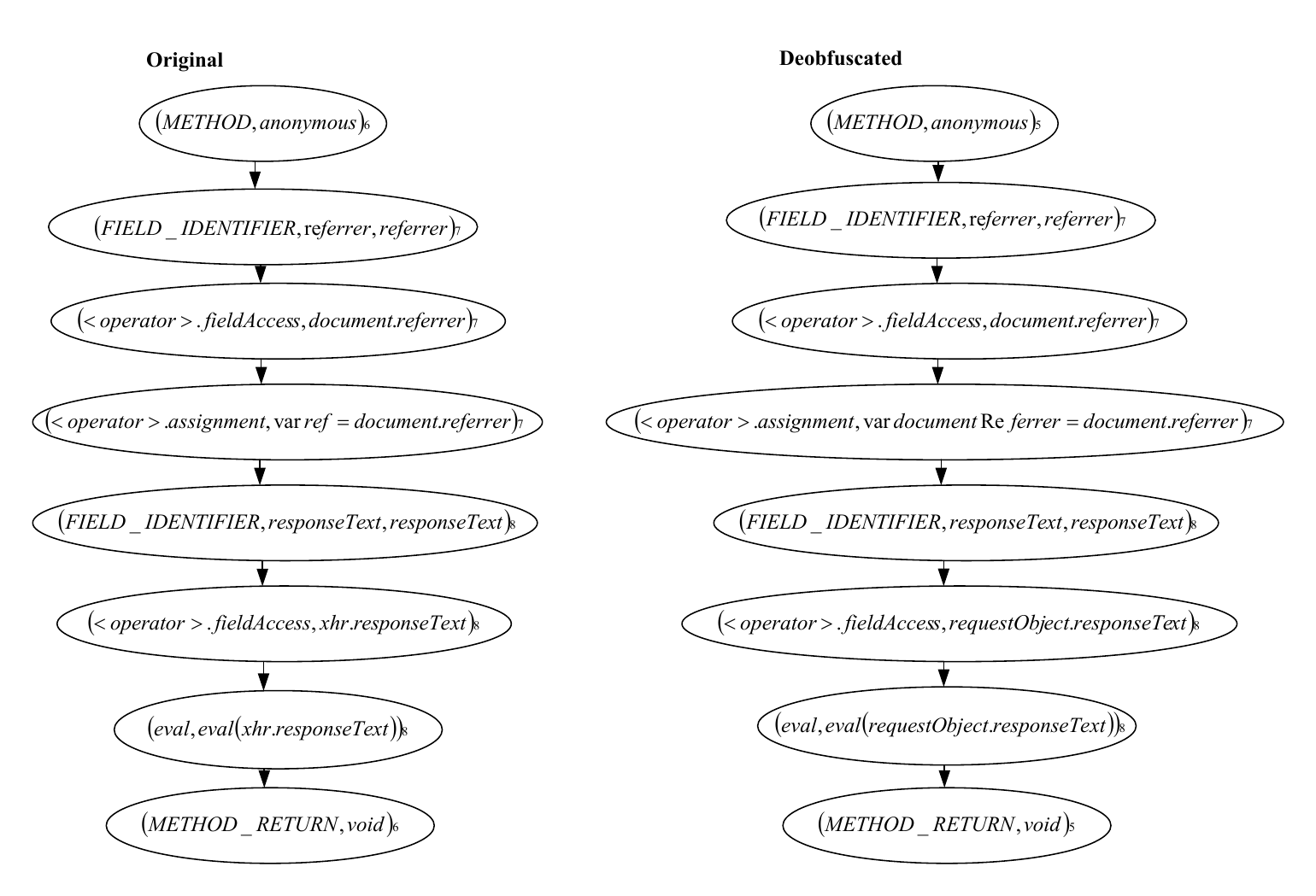}
    \caption{Control flow graphs of the original and deobfuscated versions of the sample.}
    \label{fig:cfg}
\end{figure*}

\begin{figure*}
    \centering
    \includegraphics[width=\textwidth, height=0.4\textheight, keepaspectratio]{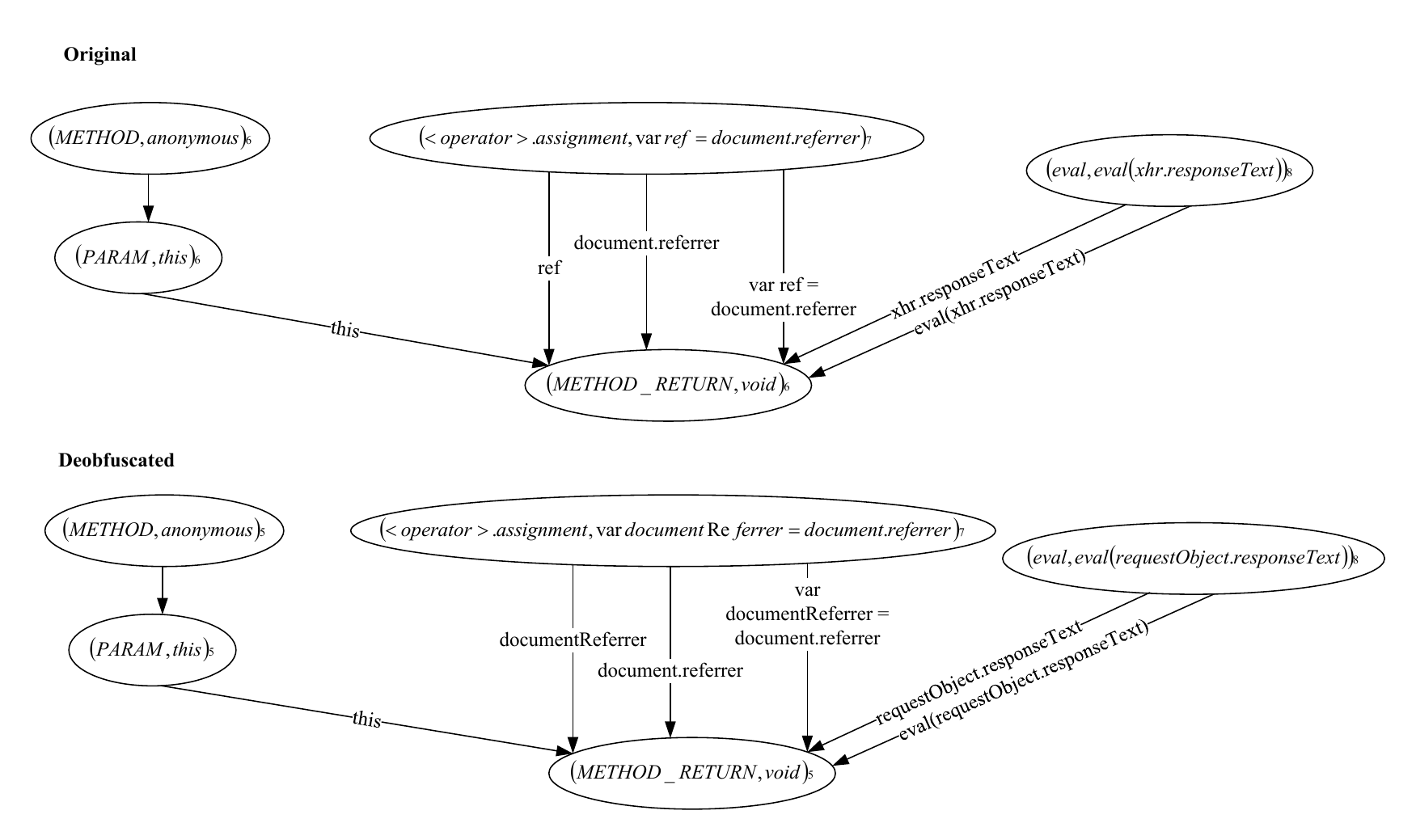}
    \caption{Data dependence graphs of the original and deobfuscated versions of the sample.}
    \label{fig:ddg}
\end{figure*}

Listing~\ref{lst:example} shows a representative sample\footnote{MD5: 000c5123cd553d14cd49d22022fcc9c5}. The original obfuscated code employed String Obfuscation (T3) and Eval Obfuscation (T15) to obscure its functionality. With the help of Joern, we generated CFGs and DDGs for both the original and deobfuscated versions, which are shown in Figure~\ref{fig:cfg} and Figure~\ref{fig:ddg}, revealing:
\begin{lstlisting}[style=JavaScript, caption={A representative sample before obfuscated.}, label={lst:example} ]
var xhr = new XMLHttpRequest();
xhr.open(
  "GET", "//lurgee.download/?zBOwv=Q10FEhkNVgQQTxNcARAJWxRMUQ5IdRJYEhUAWB1..." // truncated for brevity
);
xhr.onload = function () {
  var ref = document.referrer;
  eval(xhr.responseText);
};
xhr.send();
\end{lstlisting}
\begin{itemize}
    \item \textbf{Control Flow Analysis}: \OurTool's deobfuscated output restored the original 8-node CFG structure with 97.30\% similarity.
    \item \textbf{Data Dependency Verification}: DDG comparison showed 96.10\% preservation rate with all variable dependencies correctly maintained.
    \item \textbf{Manual Verification}: Side-by-side analysis confirmed that the core functionality of downloading and executing remote JavaScript code via \texttt{eval()} was preserved, validating semantic consistency.
\end{itemize}

\section{LLM Prompt for Evaluation}
\subsection{Deobfuscation Prompt}
\label{appendix:llm_prompt_deobfuscation}
    \begin{tcolorbox}[colback=gray!10,
        colframe=black,
        width=9cm,
        arc=2mm, 
        auto outer arc,
        title={Deobfuscation Prompt},
        breakable]	
        You are a JavaScript deobfuscation expert. Please analyze the following obfuscated JavaScript code and provide the deobfuscated version. Focus on:
        \begin{enumerate}
            \item Restoring original variable and function names where possible
            \item Simplifying complex expressions to their basic forms
            \item Converting encoded strings back to readable text
            \item Removing unnecessary complexity while preserving functionality
            \item Ensuring the output is clean, readable JavaScript code
        \end{enumerate}
        
        Obfuscated code: \textit{[OBFUSCATED\_CODE]}
        
        Please provide the deobfuscated JavaScript code and what obfuscation technology the code uses.
    \end{tcolorbox}
    
\subsection{Score Readability Prompt}
\label{appendix:llm_prompt_score}
    	\begin{tcolorbox}[colback=gray!10,
			colframe=black,
			width=9cm,
			arc=2mm, auto outer arc,
			title={Score Readability Prompt},breakable,]		
			As a security analyst, please analyze the sample according to the following steps: 1. Identification of obfuscated techniques used for raw obfuscated samples 2. Compare and evaluate the consistency of behavior between the two samples after deobfuscation 3. The readability of both versions must be rated separately (on a 0-10 point scale), format: Obfuscated sample readability score: [X] Readability score for deobfuscated samples: [Y] Please provide the final rating directly without the need for analysis process. The rating should be based on dimensions such as code structure clarity, variable naming rationality, and logical traceability.
	\end{tcolorbox}
    
\section{Anti-Analysis Technique Examples}
\label{appendix:anti_analysis_examples}

This section provides detailed code examples demonstrating how \OurTool handles various anti-analysis techniques employed by malicious JavaScript.
\begin{lstlisting}[style=JavaScript, caption={Time bomb handling example (code fragment).}, label={lst:time}]
// Original obfuscated time bomb
var _0x1a2b = new Date().getTime();
var _0x3c4d = 1640995200000;
if (_0x1a2b > _0x3c4d) { 
    eval(atob("dmFyIHBheWxvYWQ9Li4u")); 
}

// After JSIMPLIFIER processing
var currentTimestamp = new Date().getTime();
var activationDate = 1640995200000;
if (currentTimestamp > 1640995200000) {
    eval('var payload=...');
}
\end{lstlisting}
\section{Entropy Metrics: Formulations and Validation}
\label{appendix:entropy_fomulation}

\subsection{Code Text Entropy Formulation}

Code text entropy combines character-level and word-level frequency analysis:

\begin{equation}
\begin{aligned}
H_{text}(X) &= H_{char}(X) + H_{word}(X) \\
&= -\sum_{i=1}^{n} p(c_i) \log_2 p(c_i) - \sum_{j=1}^{m} p(w_j) \log_2 p(w_j)
\end{aligned}
\end{equation}

where $p(c_i)$ and $p(w_j)$ represent probabilities of character $c_i$ and word $w_j$ respectively.

\subsection{AST Entropy Component Formulations}

AST entropy incorporates four structural dimensions:

\begin{equation}
H_{node\_num}(X) = -\sum_{i=1}^{k} p(t_i) \log_2 p(t_i)
\end{equation}
\begin{equation}
H_{node\_degree}(X) = -\sum_{j=1}^{l} p(d_j) \log_2 p(d_j)
\end{equation}

where $p(t_i)$ represents node type probability and $p(d_j)$ represents node degree probability. Node depth and edge quantity entropies follow similar formulations.

The final AST entropy is computed as:
\begin{equation}
\begin{aligned}
H_{AST}(X) &= w_1 H_{node\_num} + w_2 H_{edge\_num} \\
&\quad + w_3 H_{node\_degree} + w_4 H_{node\_depth}
\end{aligned}
\end{equation}
where weights $[w_1, w_2, w_3, w_4] = [0.61, 0.79, 1.58, 1.02]$ are determined through genetic algorithm optimization to maximize discrimination between obfuscated and deobfuscated code.

\subsection{Entropy-based Evaluation Metric Validation}
We validated our entropy-based approach against traditional structural similarity metrics using 60 randomly selected samples from BenignJS. Traditional methods failed on 31 out of 60 samples (51.67\%), while AST entropy achieved stronger correlation with LLM-based readability assessment (r=0.53, p≤0.01) compared to structural similarity (r=0.25, p≥0.05), demonstrating superior robustness for evaluating deobfuscation effectiveness.
\section{Readability Analysis}
\label{appendix:readability_analysis}
\begin{figure}[htbp]
\includegraphics[width=0.5\textwidth, keepaspectratio]{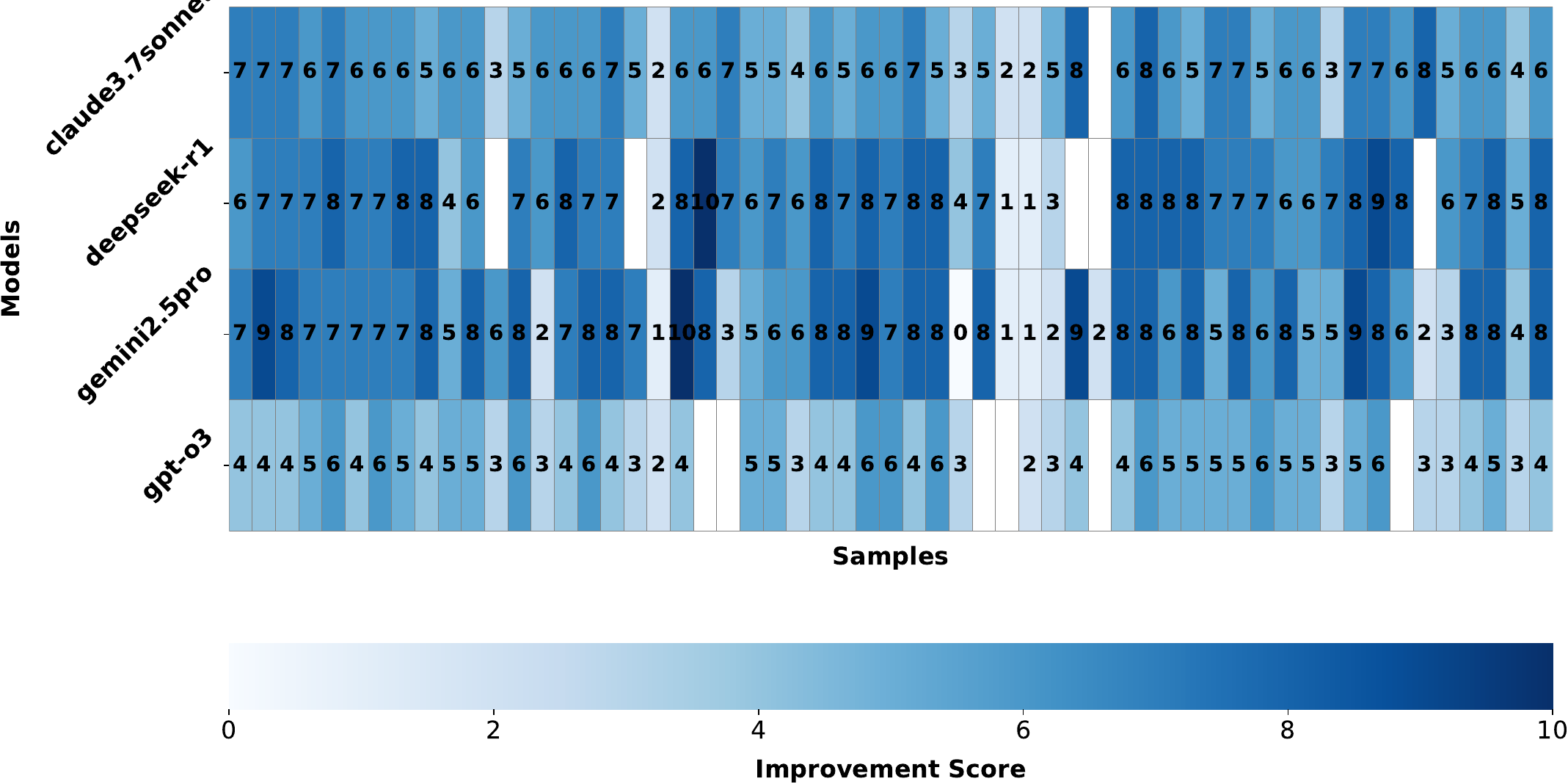}
\caption{Readability improvement scores validating complexity reduction across models and samples.}
\label{fig:improvement}
\end{figure}

Figure~\ref{fig:improvement} presents sample-level readability improvement analysis across four LLM evaluators, demonstrating consistent effectiveness of \OurTool in transforming obfuscated code into readable versions.

\section{Failure Case Example}
\label{appendix:failure_case}

This appendix presents a detailed example of semantic inconsistency introduced by LLM-based variable renaming.

\subsection{Original Code}
\begin{lstlisting}[style=JavaScript, caption={Original clean code.}, label={lst:original}]
function sumEvenSquaresOverTen(arr) {
  if (!Array.isArray(arr)) return 0;
  const filtered = arr.filter(n => n > 10 && n % 2 === 0);
  const squares = filtered.map(n => n * n);
  const sum = squares.reduce((acc, val) => acc + val, 0);
  return sum;
}
\end{lstlisting}
\onecolumn
\subsection{Obfuscated Code}
\begin{minipage}{\textwidth}
\begin{lstlisting}[style=JavaScript, caption={Obfuscated code with dynamic method access.}, label={lst:obfuscated}]
(function(){
  var _0xmethods = ['filter', 'map', 'reduce'];
  function _0xgetMethod(idx) {
    return _0xmethods[idx];
  }

  function sumEvenSquaresOverTen(arr) {
    if (!Array.isArray(arr)) return 0;
    var filtered = arr[_0xgetMethod(0)](function(n){
      return n > 10 && n % 2 === 0;
    });
    var squares = filtered[_0xgetMethod(1)](function(n){
      return n * n;
    });
    var sum = squares[_0xgetMethod(2)](function(acc, val){
      return acc + val;
    }, 0);
    return sum;
  }

  window.sumEvenSquaresOverTen = sumEvenSquaresOverTen;
})();
\end{lstlisting}
\end{minipage}
\subsection{Incorrectly Deobfuscated Code}
\begin{minipage}{\textwidth}
\begin{lstlisting}[style=JavaScript, caption={Deobfuscated code with semantic errors.}, label={lst:deobfuscated}]
'use strict';

(function () {
  var arrayMethods = ["filter", "map", "reduce"];
  function getArrayMethodByIndex(methodIndex) {
    return arrayMethods[methodIndex];
  }
  function calculateSumOfEvenSquaresAboveTen(numbersArray) {
    if (!Array.isArray(numbersArray)) {
      return 0;
    }
    var evenNumbersAboveTen = numbersArray[arrayMethods[idx]](function (number) {
      return number > 10 && number % 2 === 0;
    });
    var squaredEvenNumbers = evenNumbersAboveTen[arrayMethods[idx]](function (value) {
      return value * value;
    });
    var totalOfEvenSquaredNumbers = squaredEvenNumbers[arrayMethods[idx]](function (total, currentValue) {
      return total + currentValue;
    }, 0);
    return totalOfEvenSquaredNumbers;
  }
  window.sumEvenSquaresOverTen = calculateSumOfEvenSquaresAboveTen;
})();
\end{lstlisting}
\end{minipage}
\subsection{Error Analysis}
The deobfuscated code (Listing~\ref{lst:deobfuscated}) contains several critical semantic errors compared to the original obfuscated code (Listing~\ref{lst:obfuscated}):

\begin{itemize}
\item \textbf{Broken Variable Dependencies:} Variable \texttt{idx} is referenced but undefined, causing \texttt{ReferenceError} at runtime.
\item \textbf{Altered Control Flow:} Undefined \texttt{idx} prevents method resolution, breaking the execution flow that was correctly implemented in Listing~\ref{lst:obfuscated}.
\item \textbf{Modified API Behavior:} The code attempts to access \texttt{arrayMethods[undefined]} instead of calling the correct array methods (\texttt{filter}, \texttt{map}, \texttt{reduce}), resulting in failed API calls.
\end{itemize}

\section{User Study Results}
\label{appendix:user_study_table}
Table~\ref{tab:user_study_results} provides detailed quantitative results from our user study, showing performance metrics and subjective evaluations across three expertise levels under different analysis conditions.


\begin{table*}[htbp]
\centering
\caption{User study results: \OurTool performance and relative improvements (\%) compared to original and traditional conditions.}
\label{tab:user_study_results}
\begin{tabular}{l|ccc|ccc|ccc}
\hline
\multirow{2}{*}{\textbf{Metric}} & \multicolumn{3}{c|}{\textbf{\OurTool Performance}} & \multicolumn{3}{c|}{\textbf{vs. Original (\%)}} & \multicolumn{3}{c}{\textbf{vs. Traditional (\%)}} \\
\cline{2-10}
 & \textbf{Novice} & \textbf{Intermediate} & \textbf{Expert} & \textbf{Novice} & \textbf{Intermediate} & \textbf{Expert} & \textbf{Novice} & \textbf{Intermediate} & \textbf{Expert} \\
\hline
\rowcolor[HTML]{EFEFEF} Average Score (100) & 53.63 & 77.34 & 75.92 & +12.7 & +5.7 & +1.5 & +5.0 & +7.0 & -4.5 \\
Average Time (s) & 549.5 & 429 & 696.5 & -16.0 & -47.7 & -2.0 & -15.8 & -47.8 & -18.6 \\
\rowcolor[HTML]{EFEFEF} Readability (5) & 3.67 & 3.33 & 3.33 & +100 & +150 & +43 & +57 & +82 & +18 \\
Clarity (5) & 4 & 3.33 & 3.33 & +140 & +150 & +43 & +60 & +67 & +5 \\
\rowcolor[HTML]{EFEFEF} Logicality (5) & 4 & 3 & 3.33 & +71 & +80 & +5 & +60 & +50 & +25 \\
Difficulty (5) & 3 & 2.5 & 3.17 & +100 & +25 & +9 & +50 & +36 & +12 \\
\rowcolor[HTML]{EFEFEF} Confidence (5) & 3 & 2.5 & 4.17 & +125 & +15 & +4 & +64 & +25 & +14 \\
\hline
\end{tabular}
\end{table*}

\section{JavaScript Obfuscation Techniques Categorization and Examples}
\label{appendix:obfuscator_options}

All obfuscation examples in this Appendix use the same baseline code (Listing~\ref{lst:baseline}) to enable direct comparison of their distinct transformation effects on identical JavaScript constructs. For brevity, some obfuscation techniques are demonstrated using shorter code snippets, as indicated in the respective listings. 
\begin{center}
\begin{minipage}{0.8\textwidth}
\begin{lstlisting}[style=JavaScript, caption={Baseline code used for all obfuscation technique demonstrations.}, label={lst:baseline}]
function calculateResult() {
    var message = "Hello World";
    var number = 42;
    var isActive = true;
    console.log(message);
    return number + 1;
}
calculateResult();
\end{lstlisting}
\end{minipage}
\end{center}

For comprehensive evaluation of deobfuscation capabilities, we also used the test script shown in Listing~\ref{lst:test_code}, which encompasses various JavaScript constructs including function declarations, variable assignments, string operations, arithmetic expressions, boolean logic, and conditional statements.

\begin{center}
\begin{minipage}{0.8\textwidth}
\begin{lstlisting}[style=JavaScript, caption={Test script for deobfuscation capability evaluation.}, label={lst:test_code}]
function calculateSum(a, b) {
    var result = a + b;
    var message = "The sum is: " + result;
    var isPositive = result > 0;    
    if (isPositive) {
        console.log(message);
        return true;
    } else {
        console.log("Result is not positive");
        return false;
    }
}
var x = 10;
var y = 20;
calculateSum(x, y);
\end{lstlisting}
\end{minipage}
\end{center}

\subsection{Lexical-Level Obfuscation}

\subsubsection*{T0: Rename Identifier}
\begin{center}
\begin{minipage}{0.8\textwidth}
\begin{lstlisting}[style=JavaScript]
// Obfuscated
function _0x62ab7d() {
    var _0x3f4a8c = "Hello World";
    var _0x9b2e1f = 42;
    var _0x7c5d9a = true;
    console.log(_0x3f4a8c);
    return _0x9b2e1f + 1;
}
_0x62ab7d();
\end{lstlisting}
\end{minipage}
\end{center}

\subsubsection*{T1: Indirect Property}
\begin{center}
\begin{minipage}{0.8\textwidth}
\begin{lstlisting}[style=JavaScript]
// Obfuscated
function calculateResult() {
    var message = "Hello World";
    var number = 42;
    var isActive = true;
    console['\x6c\x6f\x67'](message);
    return number + 1;
}
calculateResult();
\end{lstlisting}
\end{minipage}
\end{center}

\subsubsection*{T2: Arithmetize}
\begin{center}
\begin{minipage}{0.8\textwidth}
\begin{lstlisting}[style=JavaScript]
// Obfuscated
function calculateResult() {
    var message = "Hello World";
    var number = 683517 ^ 683355;  // 683517 ^ 683355 = 42
    var isActive = true;
    console.log(message);
    return number + (2047 ^ 2046);  // 2047 ^ 2046 = 1
}
calculateResult();
\end{lstlisting}
\end{minipage}
\end{center}

\subsubsection*{T3: String Encode}
\begin{center}
\begin{minipage}{0.8\textwidth}
\begin{lstlisting}[style=JavaScript]
// Obfuscated
function calculateResult() {
    var message = atob("SGVsbG8gV29ybGQ=");  // Base64 for "Hello World"
    var number = 42;
    var isActive = true;
    console.log(message);
    return number + 1;
}
calculateResult();
\end{lstlisting}
\end{minipage}
\end{center}

\subsubsection*{T4: Boolean Encode}
\begin{center}
\begin{minipage}{0.8\textwidth}
\begin{lstlisting}[style=JavaScript]
// Obfuscated
function calculateResult() {
    var message = "Hello World";
    var number = 42;
    var isActive = !![];
    console.log(message);
    return number + 1;
}
calculateResult();
\end{lstlisting}
\end{minipage}
\end{center}
\vspace{0.5cm}
\subsection{Syntactic-Level Obfuscation}
\subsubsection*{T5: Expression To Function}
\begin{center}
\begin{minipage}{0.8\textwidth}
\begin{lstlisting}[style=JavaScript]
// Original
var num = 683517 ^ 683398;
// Obfuscated
var num = function (s, h) {
    return s ^ h;
}(683517, 683398);
\end{lstlisting}
\end{minipage}
\end{center}

\subsubsection*{T6: Assignment To Function}
\begin{center}
\begin{minipage}{0.8\textwidth}
\begin{lstlisting}[style=JavaScript]
// Obfuscated
function calculateResult() {
    var message;
    message = function () {
        return "Hello World";
    }();
    var number;
    number = function () {
        return 42;
    }();
    var isActive;
    isActive = function () {
        return true;
    }();
    console.log(message);
    return number + 1;
}
calculateResult();
\end{lstlisting}
\end{minipage}
\end{center}

\subsubsection*{T7: Reverse String}
\begin{center}
\begin{minipage}{0.8\textwidth}
\begin{lstlisting}[style=JavaScript]
// Obfuscated
function calculateResult() {
    var message = "dlroW olleH".split("").reverse().join("");
    var number = 42;
    var isActive = true;
    console.log(message);
    return number + 1;
}
calculateResult();
\end{lstlisting}
\end{minipage}
\end{center}

\subsubsection*{T8: AAEncode}
\begin{center}
\begin{minipage}{0.8\textwidth}
\begin{lstlisting}[style=JavaScript]
// Original
console.log("hello");
// Obfuscated (AAEncode - simplified representation)
// Note: Actual AAEncode uses special Unicode characters
// Below is a simplified ASCII representation
var a = /`m`)~-   //['_']; o=3; c=0; 
(D)={0: '_' ,wn : ((3==3) +'_') [0] ,sn :(3+ '_')[1] ,Dn:((3==3) +'_')[3] }; 
(D) [0] =((3==3) +'_') [2];(D) ['c'] = ((D)+'_') [ (3)+(3)-(0) ];
(D) ['o'] = ((D)+'_') [0];(o)=(D) ['c']+(D) ['o']+'constructor';
(D) ['_'] =(window) [o] [o];(e)='eval'; 
(D)[e]='\\'; (ooo)=('_')[2];(D) [oo]='\"';
(D) ['_'] ( (D) ['_'] (e+"console.log(\"hello\")"))(0);
// Actual AAEncode would use special Japanese/Unicode characters
\end{lstlisting}
\end{minipage}
\end{center}

\subsubsection*{T9: JJEncode}
\begin{center}
\begin{minipage}{0.8\textwidth}
\begin{lstlisting}[style=JavaScript]
// Original
alert("Hi");
// Obfuscated
hi=~[];hi={___:++hi,$$$$:(![]+"")[hi],__$:++hi,$_$_:(![]+"")[hi],_$_:++hi,$_$$:({}+"")[hi],$$_$:(hi[hi]+"")[hi],_$$:++hi,$$$_:(!""+"")[hi],$__:++hi,$_$:++hi,$$__:({}+"")[hi],$$_:++hi,$$$:++hi,$___:++hi,$__$:++hi};hi.$_=(hi.$_=hi+"")[hi.$_$]+(hi._$=hi.$_[hi.__$])+(hi.$$=(hi.$+"")[hi.__$])+((!hi)+"")[hi._$$]+(hi.__=hi.$_[hi.$$_])+(hi.$=(!""+"")[hi.__$])+(hi._=(!""+"")[hi._$_])+hi.$_[hi.$_$]+hi.__+hi._$+hi.$;hi.$$=hi.$+(!""+"")[hi._$$]+hi.__+hi._+hi.$+hi.$$;hi.$=(hi.___)[hi.$_][hi.$_];hi.$(hi.$(hi.$$+"\""+hi.$_$_+(![]+"")[hi._$_]+hi.$$$_+"\\"+hi.__$+hi.$$_+hi._$_+hi.__+"(\\\"\\"+hi.__$+hi.__$+hi.___+"\\"+hi.__$+hi.$_$+hi.__$+"\\\")\\"+hi.$$$+hi._$$+"\"")())(sojson={___:++sojson,$$$$:(![]+"")[sojson]});
\end{lstlisting}
\end{minipage}
\end{center}

\subsubsection*{T10: JSFUCK}
\begin{center}
\begin{minipage}{0.8\textwidth}
\begin{lstlisting}[style=JavaScript]
// Original
a
// Obfuscated
(![]+[])[+!![]]
\end{lstlisting}
\end{minipage}
\end{center}

\subsection{Semantic-Level Obfuscation}
\subsubsection*{T11: Arrayize Strings}
\begin{center}
\begin{minipage}{0.8\textwidth}
\begin{lstlisting}[style=JavaScript]
// Obfuscated
var _0x312g = ["Hello World"];
function calculateResult() {
    var message = _0x312g[0];
    var number = 42;
    var isActive = true;
    console.log(message);
    return number + 1;
}
calculateResult();
\end{lstlisting}
\end{minipage}
\end{center}

\subsubsection*{T12: Strings Array Encode}
\begin{center}
\begin{minipage}{0.8\textwidth}
\begin{lstlisting}[style=JavaScript]
// Obfuscated
var _0x=['65.108.101.101.102.41.94.102.123.101.109.'];
function _0xa5bdc(str,dy_key){
    dy_key=9;
    var i,k,str2='';
    k=str.split('.');
    for(i=0;i<k.length-1;i++){
        str2+=String.fromCharCode(k[i]^dy_key);
    }
    return str2;
}
function calculateResult() {
    var message = _0xa5bdc(_0x[0]);
    var number = 42;
    var isActive = true;
    console.log(message);
    return number + 1;
}
calculateResult();
\end{lstlisting}
\end{minipage}
\end{center}
\vspace{2cm}
\subsubsection*{T13: JSON Encode}
\begin{center}
\begin{minipage}{0.8\textwidth}
\begin{lstlisting}[style=JavaScript, breaklines=true, breakindent=0pt, frame=single]
// Original
var man = {"name":"tim","age":18};

// Obfuscated
var _0xeb6d9b=["114.3.41.41.43.103.104.100.108.43.51.41.43.125.96.100.43.37.3.41
.41.43.104.110.108.43.51.41.56.49.3.116."];
function _0xf72b(str,dy_key){
    dy_key=9;
    var i,k,str2="";
    k=str.split(".");
    for(i=0;i<k.length-1;i++){
        str2+=String.fromCharCode(k[i]^dy_key);
    }
    return str2;
}
var man=JSON.parse(_0xf72b(_0xeb6d9b[0]));
\end{lstlisting}
\end{minipage}
\end{center}

\subsubsection*{T14: Regexp Encode}
\begin{center}
\begin{minipage}{0.8\textwidth}
\begin{lstlisting}[style=JavaScript, breaklines=true, breakindent=0pt, frame=single]
// Original
var r =/regexp test/g;
// Obfuscated
var _0x796d=["123.108.110.108.113.121.41.125.108.122.125.","110."];
function _0xcca(str,dy_key){
    dy_key=9;
    var i,k,str2="";
    k=str.split(".");
    for(i=0;i<k.length-1;i++){
        str2+=String.fromCharCode(k[i]^dy_key);
    }
    return str2;
}
var r=new RegExp(_0xcca(_0x796d[0]),_0xcca(_0x796d[1]));
\end{lstlisting}
\end{minipage}
\end{center}

\subsubsection*{T15: Eval}
\begin{center}
\begin{minipage}{0.8\textwidth}
\begin{lstlisting}[style=JavaScript, breaklines=true, breakindent=0pt, frame=single]
// Obfuscated
eval(
  (function (p, a, c, k, e, d) {
    e = function (c) {
      return (
        (c < a ? "" : e(parseInt(c / a))) +
        ((c = c % a) > 35 ? String.fromCharCode(c + 29) : c.toString(36))
      );
    };
    if (!"".replace(/^/, String)) {
      while (c--) d[e(c)] = k[c] || e(c);
      k = [function (e) { return d[e];},];
      e = function () {
        return "\\w+";
      };
      c = 1;
    }
    while (c--)
      if (k[c]) p = p.replace(new RegExp("\\b" + e(c) + "\\b", "g"), k[c]);
    return p;
  })(
    '5 0(){4 1="3 6";4 2=7;4 8=9;a.b(1);c 2+d;}0();',62,14,"calculateResult|message|number|Hello|var|function|World|42|isActive|true|console|log|return|1".split("|"),0,{}));
\end{lstlisting}
\end{minipage}
\end{center}

\subsubsection*{T16: Flattern Control Flow}
\begin{center}
\begin{minipage}{0.8\textwidth}
\begin{lstlisting}[style=JavaScript]
// Obfuscated
function calculateResult() {
    var _array = "1|0|3|2|4".split("|"),
    _index = 0;
    while (!![]) {
        switch (+_array[_index++]) {
        case 0:
            var number = 42;
            continue;
        case 1:
            var message = "Hello World";
            continue;
        case 2:
            console.log(message);
            continue;
        case 3:
            var isActive = true;
            continue;
        case 4:
            return number + 1;
            continue;
        }
        break;
    }
}
calculateResult();
\end{lstlisting}
\end{minipage}
\end{center}

\subsubsection*{T17: Insert Dead Code}
\begin{center}
\begin{minipage}{0.8\textwidth}
\begin{lstlisting}[style=JavaScript, breaklines=true]
// Obfuscated
function calculateResult() {
    var unusedVar = Math.random() * 1000;
    if (false) {
        console.log("This will never execute");
        return unusedVar;
    }
    var message = "Hello World";
    var number = 42;
    var isActive = true;
    console.log(message);
    
    if (1 === 0) {
        return number * unusedVar;
    }
    return number + 1;
    // Unreachable dead code
    function deadFunction() {
        return unusedVar + 42;
    }
}
calculateResult();
\end{lstlisting}
\end{minipage}
\end{center}

\subsection{Multi-layered Obfuscation}
\subsubsection*{T18: OB Obfuscation}
\begin{center}
\begin{minipage}{0.8\textwidth}
\begin{lstlisting}[style=JavaScript]
// Obfuscated
var _0x30bb = ['log', 'Hello\x20World', '42', 'true', 'calculateResult'];
(function (_0x38d89d, _0x30bbb2) {
    var _0xae0a32 = function (_0x2e4e9d) {
        while (--_0x2e4e9d) {
            _0x38d89d['push'](_0x38d89d['shift']());
        }
    };
    _0xae0a32(++_0x30bbb2);
}(_0x30bb, 0x153));

var _0xae0a = function (_0x38d89d, _0x30bbb2) {
    _0x38d89d = _0x38d89d - 0x0;
    var _0xae0a32 = _0x30bb[_0x38d89d];
    return _0xae0a32;
};
function calculateResult() {
    var message = _0xae0a('0x1');
    var number = parseInt(_0xae0a('0x2'));
    var isActive = _0xae0a('0x3') === 'true';
    console[_0xae0a('0x0')](message);
    return number + 1;
}
window[_0xae0a('0x4')]();
\end{lstlisting}
\end{minipage}
\end{center}

\subsubsection*{T19: LLM-based Obfuscation}
Recent research by Palo Alto Networks Unit 42~\cite{unit42:llm_obfuscation} demonstrated that LLMs can rewrite malware with transformations that appear more natural than traditional obfuscation tools. Inspired by this work, we developed T19 to simulate LLM-generated obfuscation patterns using natural language prompts. Unlike predefined transformations, this technique combines multiple subtle changes including dead code insertion, variable renaming, string manipulation, and alternative algorithmic implementations that collectively obscure the original intent while maintaining semantic equivalence.
\begin{center}
\begin{minipage}{0.8\textwidth}
\begin{lstlisting}[style=JavaScript]
// Obfuscated (LLM-generated with multiple techniques)
if (false) { console.log("Dead code"); }
var unusedVar = 999;
function computeOutput() {
    // String splitting
    var msgParts = ["H", "e", "l", "l", "o", " ", "W", "o", "r", "l", "d"];
    var textContent = msgParts.join('');
    // Alternative implementation
    var numericValue = (50 - 8); // Gets 42
    var statusFlag = !false; // Gets true
    console.log(textContent);
    return numericValue + 1;
}
computeOutput();
\end{lstlisting}
\end{minipage}
\end{center}

\begin{tcolorbox}[
    colback=gray!10,
    colframe=black,
    width=\textwidth,
    arc=2mm, 
    auto outer arc,
    title={LLM Obfuscation Prompt},
    breakable,
    enhanced,
    left=1mm,
    right=1mm,
    top=1mm,
    bottom=1mm
]
\small
\textbf{System Role:} You are a JavaScript obfuscation tool specialized in applying specific transformation techniques.

\textbf{Available Obfuscation Techniques:}
\begin{itemize}
    \item \textbf{dead\_code\_insert}: Insert non-functional code while preserving original functionality
    \item \textbf{var\_rename}: Rename identifiers to random or meaningful names
    \item \textbf{string\_splitting}: Convert string literals to character arrays
    \item \textbf{alternative\_reimplementation}: Rewrite functions using different algorithms
\end{itemize}

\textbf{Output Format:}
\begin{verbatim}
[START ORIGINAL JAVASCRIPT]
```javascript
// original code here
[END ORIGINAL JAVASCRIPT] [START EXPLANATION]
// transformation explanation
[END EXPLANATION] [START OBFUSCATED JAVASCRIPT]
// obfuscated code here
[END OBFUSCATED JAVASCRIPT]
\end{verbatim} \textbf{Task:} Apply the specified obfuscation technique to the provided JavaScript code while maintaining identical functionality.
\end{tcolorbox}

\section{User Study Questionnaire}
\label{appendix:questionnaire}

\subsection{Objective Analysis Tasks (100 points total)}

\subsubsection{Code Understanding Tasks (40 points)}
This section evaluates participants' ability to comprehend code structure and functionality.

\textbf{T1: Code Functionality Identification (10 points)}
Which of the following main functions does this code contain? (Multiple choice)
\begin{itemize}
    \item[A.] Data collection and theft: collecting user input, system information, browser data, file content, etc.
    \item[B.] Network communication and control: initiating network requests (HTTP/WebSocket), downloading payloads, establishing C2 channels, transmitting data back.
    \item[C.] Payload decryption and execution: decrypting encrypted/encoded strings or payloads and preparing or executing them.
    \item[D.] Persistence and residence: attempting to write files, registry, scheduled tasks, etc. to survive after restart.
    \item[E.] Defense evasion: executing other non-obfuscation anti-analysis behaviors (e.g., privilege escalation, log clearing, disabling security software).
\end{itemize}

\textbf{T2: Key Variable Identification (10 points)}
Which of the following are key variables in the code? (Multiple choice, maximum 3 selections)
[Options provided based on specific sample]

\textbf{T3: Execution Flow Judgment (10 points)}
What is the execution order of the code? Please arrange the following steps in correct order. (Drag and drop sorting)
[4-5 execution steps provided based on sample]

\textbf{T4: Conditional Branch Analysis (10 points)}
Under what conditions will the code execute specific branches? (Single choice)
\begin{itemize}
    \item[A.] When variable X is greater than a certain value
    \item[B.] When a specific environment is detected
    \item[C.] When specific input is received
    \item[D.] When time conditions are met
    \item[E.] Unconditional execution
\end{itemize}

\subsubsection{Threat Analysis Tasks (40 points)}
This section assesses participants' capability to identify malicious behaviors and security threats.

\textbf{T5: Malicious Behavior Type Identification (15 points)}
What malicious behaviors might this code execute? (Multiple choice)
\begin{itemize}
    \item[A.] Steal user credentials
    \item[B.] Collect system information
    \item[C.] Establish remote connections
    \item[D.] Modify system files
    \item[E.] Monitor user behavior
    \item[F.] Redirect network traffic
    \item[G.] Execute unauthorized code
    \item[H.] Propagate malicious payloads
\end{itemize}

\textbf{T6: Attack Target Identification (10 points)}
What is the ultimate attack target of this code? (Single choice)
\begin{itemize}
    \item[A.] Steal sensitive information
    \item[B.] Gain system control
    \item[C.] Hijack user sessions/traffic
    \item[D.] Achieve persistent residence
    \item[E.] Destroy system or data
\end{itemize}

\textbf{T7: IOC (Indicators of Compromise) Extraction (15 points)}
Please extract all observable threat indicators (IOCs) from the code:
\begin{itemize}
    \item Network-related indicators (5 points): URLs/domains, IP addresses, port numbers
    \item File/path indicators (3 points): file paths, file names
    \item System/API indicators (4 points): API calls, system commands, registry keys
    \item Encryption/encoding indicators (3 points): encryption keys/salt values, encoded strings, algorithm identifiers
\end{itemize}

\subsubsection{Technical Analysis Tasks (20 points)}
This section measures participants' technical analysis skills and code complexity assessment.

\textbf{T8: Code Feature Identification (15 points)}
What features does this code have? (Multiple choice)
\begin{itemize}
    \item[1.] Uses complex string operations
    \item[2.] Contains dynamic code execution
    \item[3.] Has conditional branch logic
    \item[4.] Uses array or object operations
    \item[5.] Contains network communication code
    \item[6.] Has loop or iteration structures
    \item[7.] Uses built-in API calls
    \item[8.] Contains error handling mechanisms
    \item[9.] Has time or delay-related operations
    \item[10.] Uses regular expressions
\end{itemize}

\textbf{T9: Code Complexity Assessment (5 points)}
From an analysis perspective, how complex is this code? (Single choice)
\begin{itemize}
    \item[1.] Simple (clear logic, easy to understand)
    \item[2.] Medium (requires careful analysis to understand)
    \item[3.] Complex (requires in-depth analysis of multiple layers)
    \item[4.] Extremely complex (requires professional tools and experience)
\end{itemize}

\subsection{Subjective Evaluation Tasks}

\subsubsection{Readability Assessment (5-point Likert scale)}

\begin{enumerate}
    \item \textbf{Overall code readability:} How readable is the code? (1 = Completely unreadable, 5 = Very readable)
    \item \textbf{Clarity of variable and function names:} How clear are the variable and function names? (1 = Completely unclear, 5 = Very clear)
    \item \textbf{Logical structure of the code:} How logical and well-structured is the code? (1 = Completely chaotic, 5 = Very clear and logical)
    \item \textbf{Difficulty of analyzing the code:} How difficult was it to analyze this code? (1 = Very difficult, 5 = Very easy)
    \item \textbf{Compared to the original obfuscated code, how does the analysis difficulty of the current version compare?} (Significantly harder / Slightly harder / Same / Slightly easier / Significantly easier)
    \item \textbf{Confidence in your analysis results:} How confident are you in the correctness of your analysis? (1 = Not confident at all, 5 = Very confident)
\end{enumerate}
\twocolumn
\section{Artifact Appendix}
\label{appendix:artifact}

 

\subsection{Description \& Requirements}




\subsubsection{How to access}


The artifact is archived on Zenodo with DOI \href{https://doi.org/10.5281/zenodo.17531662}{10.5281/zenodo.17531662} for long-term preservation and reproducibility.

The artifact is publicly available and can be accessed directly through the DOI link above. All materials, including source code, datasets, and documentation, are provided for full community transparency and reuse.

\subsubsection{Hardware dependencies}

None. JSIMPLIFIER and its evaluation experiments impose minimal hardware requirements, allowing flexible deployment on common computing platforms. Our experimental setup utilized an 8-core AMD EPYC 7742 processor with 16 GB of RAM running Windows 11; however, the tool is not restricted to specific hardware and can operate effectively on comparable or even modest systems depending on workload scale.

\subsubsection{Software dependencies}

\begin{itemize}
    \item Core Runtime:
    \begin{itemize}
    \item  Node.js v20.9.0
    \item  Python 3.11.8
    \end{itemize}
    \item LLM Integration (Optional): API access to: OpenAI GPT, Google Gemini
    \item Additional Dependencies:
    \begin{itemize}
    \item  All Python packages listed in \texttt{requirements.txt}
    \item  All Node.js packages listed in \texttt{package.json}
    \end{itemize}
\end{itemize}

\subsubsection{Benchmarks}

\begin{itemize}
    \item Primary Datasets:
    \begin{itemize}
    \item MalJS: 23,212 real-world obfuscated malicious JavaScript samples (391.78 KB average size)
    \item BenignJS: 21,209 real-world obfuscated benign JavaScript samples (41.40 KB average size)
    \end{itemize}
\end{itemize}

\subsection{Artifact Installation \& Configuration}

\begin{enumerate}
    \item \textbf{Download and extract the artifact:}

    {\small
    \begin{verbatim}
# Download from Zenodo
# Extract the downloaded archive
unzip jsimplifier.zip
cd jsimplifier
    \end{verbatim}
    }

    \item \textbf{Install Node.js dependencies:}
    
    {\small
    \begin{verbatim}
npm install
    \end{verbatim}
    }
    \item \textbf{Install Python dependencies:}
    
    {\small
    \begin{verbatim}
pip install -r requirements.txt
    \end{verbatim}
    }
    \item \textbf{Configure LLM API access (optional):}
    
    Set environment variables for API access:
    
    {\small
    \begin{verbatim}
export OPENAI_API_KEY="your-openai-key"
export GEMINI_API_KEY="your-gemini-key"
export BASE_URL="your-baseurl"
    \end{verbatim}
    }
    Or provide API keys directly via command line parameters during execution.

\end{enumerate}

\subsection{Experiment Workflow}

The experimental evaluation is designed for practical artifact assessment with flexible scaling options to accommodate different evaluation time constraints:

\textbf{Stage 1}: Deobfuscation Capability Demonstration - Interactive testing of individual obfuscation techniques with provided sample cases

\textbf{Stage 2}: Syntax Validation - Automated verification that deobfuscated output is syntactically correct JavaScript  

\textbf{Stage 3}: Scalable Effectiveness Validation - Process real-world samples with configurable dataset size (from quick demos to full-scale evaluation)

Each stage provides clear, verifiable outputs with time estimates ranging from minutes (demo mode) to days (full evaluation).


\subsection{Major Claims}

\begin{itemize}
    \item \textbf{(C1)}: JSIMPLIFIER successfully handles all 20 categorized obfuscation techniques. This is demonstrated by experiment (E1) using provided sample cases for each technique.
    
    \item \textbf{(C2)}: JSIMPLIFIER produces syntactically valid JavaScript output. This is validated by experiment (E2) using automated syntax checking.
    
    \item \textbf{(C3)}: JSIMPLIFIER demonstrates robust processing capability on real-world samples at scale. This is shown by experiment (E3) with configurable dataset sizes to balance evaluation thoroughness with time constraints.
\end{itemize}



\subsection{Evaluation}

\subsubsection{Experiment (E1): Deobfuscation Capability Demonstration}
\textbf{[15 human-minutes + 10 compute-minutes]}: Validate JSIMPLIFIER's capability to handle different obfuscation techniques using provided sample cases.

\textit{[Preparation]}
    {\small
    \begin{verbatim}
cd experiments/E1
    \end{verbatim}
    }

\textit{[Execution]}
{\small
\begin{verbatim}
# Test individual techniques
npm start deobfuscate -- --model=... \
  --apiKey=... --baseURL=... \
  experiments/E1/obfuscated_samples/T0_Rename.js
npm start deobfuscate -- --model=none \
  experiments/E1/obfuscated_samples/T1_Indirect.js
# ... or run all samples
python .\run_technique_demos.py --model ... \
  --api-key ... --base-url ...
\end{verbatim}
}

\textit{[Results]}

Expected: Each sample produces readable, deobfuscated output. Results saved in \texttt{results/} with side-by-side comparisons.

\subsubsection{Experiment (E2): Syntax Validation}
\textbf{[10 human-minutes + 5 compute-minutes]}: Verify that deobfuscated output is syntactically valid JavaScript.

\textit{[Preparation]}
{\small
\begin{verbatim}
cd experiments/E2
\end{verbatim}
}

\textit{[Execution]}
{\small
\begin{verbatim}
# Validate syntax of deobfuscated files
python validate_syntax.py \
  ../E1/results/
\end{verbatim}
}

\textit{[Results]}

Expected: 100\% syntax validation success. Output shows parsing results for each deobfuscated file.

\subsubsection{Experiment (E3): Scalable Effectiveness Validation}
\textbf{Time Options: [Quick: 30 minutes] [Standard: 4 hours] [Full: ~8 days single-core, ~1-2 days multi-core]}: Demonstrate tool robustness on real-world malicious JavaScript samples with flexible scale and parallelization options.

\textit{[Preparation]}
{\small
\begin{verbatim}
tar -zxvf dataset/MalJS.tar.gz
cd experiments/E3

\end{verbatim}
}

\textit{[Execution Options]}

\textbf{Quick Demo (30 minutes):}
{\small
\begin{verbatim}
# Process 50 representative samples
python run_validation.py \
  --dataset quick_demo \
  --workers 1
\end{verbatim}
}

\textbf{Standard Evaluation (4 hours):}
{\small
\begin{verbatim}
# Process 500 samples with multi-core
python run_validation.py \
  --dataset standard_eval \
  --workers 4
\end{verbatim}
}

\textbf{Full Dataset Processing:}
{\small
\begin{verbatim}
# Single-core (~8 days): 
python run_validation.py \
  --dataset full_dataset \
  --workers 1

# Multi-core (1-2 days):
python run_validation.py \
  --dataset full_dataset \
  --workers 8
# Adjust --workers based on 
# available cores
\end{verbatim}
}

\textit{[Results]}

Expected: Entropy reduction validation demonstrating deobfuscation effectiveness. The script generates box plots comparing entropy values before and after deobfuscation, showing significant complexity reduction across processed samples. See \path{output_validation_quick_demo\entropy_comparison_boxplot.png} in the root directory for visual confirmation of entropy reduction results

\subsection{Notes}
\textbf{Dataset and Cost Considerations:}
The complete MalJS dataset (23,212 samples) is provided for full reproducibility. LLM-based experiments require API access and incur approximately \$480-500 for full evaluation using GPT-4o-mini, though all core claims can be validated without LLM processing using AST-only mode. Scaled subsets maintain statistical representativeness while reducing both evaluation time and costs.

\textbf{Performance Characteristics:}
JSIMPLIFIER processes samples at an average rate of ~28 seconds per file (excluding LLM processing), with memory requirements scaling linearly with dataset size (8GB sufficient for standard evaluation). Full dataset processing requires approximately 8 days on single-core systems or 1-2 days with 8+ cores, as performance is primarily CPU-bound. Network connectivity is only required for optional LLM-based enhancements.

\textbf{Evaluation Flexibility:}
The artifact supports three evaluation scales: quick validation (30 minutes) verifies core functionality with representative samples, standard evaluation (4 hours) provides statistical confidence through multi-core processing, and full reproduction (1-3 days depending on available cores) processes the complete dataset. LLM processing adds significant overhead and is not recommended for full dataset evaluation due to cost and time constraints.

\end{CJK*}
\end{document}